\documentclass[10pt]{article}
\pdfoutput=1

\usepackage{rotating}
\usepackage{amssymb}
\usepackage{amsmath}
\usepackage{amsthm}
\usepackage{mathtools}
\usepackage{float}
\usepackage{caption}
\usepackage{subcaption}
\usepackage[usenames,dvipsnames]{xcolor}
\usepackage{epsfig}
\usepackage{dcolumn}
\usepackage{tikz}
\usepackage{tikz-cd}
\usetikzlibrary{math}
\usetikzlibrary{fpu}
\usetikzlibrary{shapes.geometric}
\usetikzlibrary{decorations.markings,snakes}
\usetikzlibrary{shapes.misc}
\usetikzlibrary{calc}
\usepackage{mathtools}
\usepackage{upgreek}
\usepackage{setspace}
\usepackage{enumitem}
\usepackage{array,multirow,bigdelim,arydshln}
\usepackage{appendix}
\usepackage[export]{adjustbox}
\usepackage{xparse}
\usepackage[utf8]{inputenc}
\usepackage{microtype}
\usepackage{bm}
\usepackage{bbm}
\usepackage{braket}
\usepackage{dsfont}
\usepackage{nccmath}
\usepackage{scalerel}
\usepackage[dvipsnames]{xcolor}
\definecolor{rindou1}{rgb}{0.4431,0.2862,0.7960}
\definecolor{rindou2}{rgb}{0.0078,0.1215,0.4392}
\definecolor{lapis}{rgb}{0.0.0470,0.2941,0.5568}
\definecolor{emerald}{rgb}{0.31, 0.78, 0.47}
\definecolor{pinegreen}{rgb}{0.0, 0.47, 0.44}
\definecolor{jade}{rgb}{0.0, 0.66, 0.42}
\definecolor{teal}{rgb}{0.0, 0.5, 0.5}
\usepackage{jheppub}

\usepackage{hyperref}
\hypersetup{
	colorlinks=true,
	urlcolor=Violet,
	linkcolor=Violet,
	citecolor=Violet,
	bookmarks=true,
	pdfauthor={},
	pdftitle={},
	pdfdisplaydoctitle=true,
	pdfstartview=FitH,
	backref=false,
	pagebackref=false
}
\definecolor{orcidlogocol}{HTML}{A6CE39}
\tikzset{
	orcidlogo/.pic={
		\fill[orcidlogocol] svg{M256,128c0,70.7-57.3,128-128,128C57.3,256,0,198.7,0,128C0,57.3,57.3,0,128,0C198.7,0,256,57.3,256,128z};
		\fill[white] svg{M86.3,186.2H70.9V79.1h15.4v48.4V186.2z}
		svg{M108.9,79.1h41.6c39.6,0,57,28.3,57,53.6c0,27.5-21.5,53.6-56.8,53.6h-41.8V79.1z M124.3,172.4h24.5c34.9,0,42.9-26.5,42.9-39.7c0-21.5-13.7-39.7-43.7-39.7h-23.7V172.4z}
		svg{M88.7,56.8c0,5.5-4.5,10.1-10.1,10.1c-5.6,0-10.1-4.6-10.1-10.1c0-5.6,4.5-10.1,10.1-10.1C84.2,46.7,88.7,51.3,88.7,56.8z};
	}
}

\newcommand\orcidicon[1]{\raisebox{.05em}{\href{https://orcid.org/#1}{\mbox{\scalerel*{
					\begin{tikzpicture}[yscale=-1,transform shape]
						\pic{orcidlogo};
					\end{tikzpicture}
				}{|}}}}}

\allowdisplaybreaks
\graphicspath{{figures/}}

\newcommand{\dd}{\mathrm{d}}

\title{On universal splittings of tree-level particle and string scattering amplitudes
}

\author[a,b]{Qu Cao,} \author[a,c]{Jin Dong,}\author[a,d,e,f]{Song He,} \author[a]{Canxin Shi,}\author[a,c,d]{Fanky Zhu}

\emailAdd{qucao@zju.edu.cn}\emailAdd{dongjin@itp.ac.cn}\emailAdd{songhe@itp.ac.cn}\emailAdd{shicanxin@itp.ac.cn}\emailAdd{zhufan22@mails.ucas.ac.cn}

\affiliation[a]{CAS Key Laboratory of Theoretical Physics, Institute of Theoretical Physics, Chinese Academy of Sciences, Beijing 100190, China}
\affiliation[b]{Zhejiang Institute of Modern Physics, Department of Physics, Zhejiang University, Hangzhou 310027, China}
\affiliation[c]{School of Physical Sciences, University of Chinese Academy of Sciences, Beijing 100049, China}
\affiliation[d]{School of Fundamental Physics and Mathematical Sciences,
Hangzhou Institute for Advanced Study}
\affiliation[e]{International Centre for Theoretical Physics Asia-Pacific, UCAS, Hangzhou 310024, China}

\affiliation[f]{Peng Huanwu Center for Fundamental Theory, Hefei 230026, China}

\abstract{In this paper, we study the newly discovered universal splitting behavior for tree-level scattering amplitudes of particles and strings~\cite{Cao:2024gln}: when a set of Mandelstam variables (and Lorentz products involving polarizations for gluons/gravitons) vanish, the $n$-point amplitude factorizes as the product of two lower-point {\it currents} with $n{+}3$ external legs in total. We refer to any such subspace of the kinematic space of $n$ massless momenta as ``2-split kinematics", where the scattering potential for string amplitudes and the corresponding scattering equations for particle amplitudes nicely split into two parts. Based on these, we provide a systematic and detailed study of the splitting behavior for essentially all ingredients which appear as integrands for open- and closed-string amplitudes as well as Cachazo-He-Yuan (CHY) formulas, including Parke-Taylor factors, correlators in superstring and bosonic string theories, and CHY integrands for a variety of amplitudes of scalars, gluons and gravitons. These results then immediately lead to the splitting behavior of string and particle amplitudes in a wide range of theories, including bi-adjoint $\phi^3$ (with string extension known as $Z$ and $J$ integrals), non-linear sigma model, Dirac-Born-Infeld, the special Galileon, {\it etc.}, as well as Yang-Mills and Einstein gravity (with bosonic and superstring extensions). Our results imply and extend some other factorization behavior of tree amplitudes considered recently, including smooth splittings~\cite{Cachazo:2021wsz} and factorizations near zeros~\cite{Arkani-Hamed:2023swr}, to all these theories. A special case of splitting also yields soft theorems for gluons/gravitons as well as analogous soft behavior for Goldstone particles near their Adler zeros. }

\begin{document}
\maketitle
\setcounter{page}{2}

\pagebreak
\section{Introduction and Review}

\begin{figure*}[!htbp]
    \centering
\begin{tikzpicture}
	\shade[ball color = gray, opacity=0.7] (0,0) circle (0.6);
	\node (p) at (0,0) {};
	\draw[thick,decorate,decoration=snake, segment length=8pt, segment amplitude=1.5pt] (45:0.6)--++(45:0.5) ;
	\draw[thick,decorate,decoration=snake, segment length=8pt, segment amplitude=1.5pt] (-45:0.6)--++(-45:0.5) ;
	\draw[thick,decorate,decoration=snake, segment length=8pt, segment amplitude=1.5pt] (135:0.6)--++(135:0.5) ;
	\draw[thick,decorate,decoration=snake, segment length=8pt, segment amplitude=1.5pt] (-135:0.6)--++(-135:0.5) ;
	\draw (135:1.2) node[left=-2pt]{\scriptsize $1$};
	\draw (-135:1.2) node[left=-2pt]{\scriptsize $n$};
	\draw (90:0.6) node[above=0pt]{\scriptsize $\ldots$};
	\draw (270:0.6) node[below=0pt]{\scriptsize $\ldots$};

	\node at (2,0) {\scriptsize	$\xrightarrow[]{s_{a,b}=0,a\in A,b\in B}$};

	\shade[ball color = gray, opacity=0.7] (5,0) circle (0.6);
	\node (p1) at (5,0) {};
	\draw[thick,decorate,decoration=snake, segment length=8pt, segment amplitude=1.5pt] ($(p1)+(45:0.6)$)--++(45:0.5) ;
	\draw[thick,decorate,decoration=snake, segment length=8pt, segment amplitude=1.5pt] ($(p1)+(-45:0.6)$)--++(-45:0.5) ;
	\draw[thick] ($(p1)+(135:0.6)$)--++(135:0.5) ;
	\draw[thick] ($(p1)+(-135:0.6)$)--++(-135:0.5) ;
	\draw[thick,double] ($(p1)+(180:0.6)$)--++(180:0.4) ;
	\draw ($(p1)+(180:1.1)$) node[left=-2pt]{\scriptsize $\kappa$};
	\draw ($(p1)+(135:1.2)$) node[left=-2pt]{\scriptsize $i$};
	\draw ($(p1)+(-135:1.2)$) node[left=-2pt]{\scriptsize $j$};
	\draw ($(p1)+(90:0.6)$) node[above=0pt]{\scriptsize $\ldots$};
	\draw ($(p1)+(270:0.6)$) node[below=0pt]{\scriptsize $\ldots$};

    \draw[thick,decorate,decoration=snake, segment length=8pt, segment amplitude=1.5pt] ($(p1)+(120:0.6)$)--++(120:0.45) ;
    \draw[thick,decorate,decoration=snake, segment length=8pt, segment amplitude=1.5pt] ($(p1)+(-120:0.6)$)--++(-120:0.45) ;
    \draw ($(p1)+(120:1.1)$) node[above=-2pt]{\scriptsize $i+1$};
    \draw ($(p1)+(-120:1.1)$) node[below=-2pt]{\scriptsize $j-1$};

	\node at (6.5,0) {\scriptsize	$\times$};

	\shade[ball color = gray, opacity=0.7] (8,0) circle (0.6);
	\node (p2) at (8,0) {};
	\draw[thick,decorate,decoration=snake, segment length=8pt, segment amplitude=1.5pt] ($(p2)+(45:0.6)$)--++(45:0.5) ;
	\draw[thick,decorate,decoration=snake, segment length=8pt, segment amplitude=1.5pt] ($(p2)+(-45:0.6)$)--++(-45:0.5) ;
	\draw[thick,decorate,decoration=snake, segment length=8pt, segment amplitude=1.5pt] ($(p2)+(135:0.6)$)--++(135:0.5) ;
	\draw[thick,decorate,decoration=snake, segment length=8pt, segment amplitude=1.5pt] ($(p2)+(-135:0.6)$)--++(-135:0.5) ;
	\draw[thick,double,decorate,decoration=snake, segment length=8pt, segment amplitude=1.5pt] ($(p2)+(0:0.6)$)--++(0:0.4) ;
	\draw ($(p2)+(0:1.1)$) node[right=-2pt]{\scriptsize $\kappa^\prime$};
	\draw ($(p2)+(135:1.2)$) node[left=-2pt]{\scriptsize $j$};
	\draw ($(p2)+(-135:1.2)$) node[left=-2pt]{\scriptsize $i$};
	\draw ($(p2)+(45:1.2)$) node[right=-2pt]{\scriptsize $n-1$};
	\draw ($(p2)+(-45:1.2)$) node[right=-2pt]{\scriptsize $1$};
	\draw ($(p2)+(90:0.6)$) node[above=0pt]{\scriptsize $\ldots$};
	\draw ($(p2)+(270:0.6)$) node[below=0pt]{\scriptsize $\ldots$};

    \draw[thick,decorate,decoration=snake, segment length=8pt, segment amplitude=1.5pt] ($(p2)+(120:0.6)$)--++(120:0.45) ;
    \draw[thick,decorate,decoration=snake, segment length=8pt, segment amplitude=1.5pt] ($(p2)+(-120:0.6)$)--++(-120:0.45) ;
    \draw ($(p2)+(120:1.1)$) node[above=-2pt]{\scriptsize $j+1$};
    \draw ($(p2)+(-120:1.1)$) node[below=-2pt]{\scriptsize $i-1$};

	\node at (6.3,-1.5) {\tiny $ s_{a,n}=0,a\in A\backslash\{m\}$};
	\node at (5,-1.5) {\scriptsize $\Bigg\downarrow$};

	\node at (5,-3) {\scriptsize $ \times$};
	\shade[ball color = gray, opacity=0.7] (6.5,-3) circle (0.6);
	\node (q1) at (6.5,-3) {};
	\draw[thick,double] ($(q1)+(45:0.6)$)--++(45:0.5) ;
	\draw[thick] ($(q1)+(-45:0.6)$)--++(-45:0.5) ;
	\draw[thick] ($(q1)+(135:0.6)$)--++(135:0.5) ;
	\draw[thick,double] ($(q1)+(-135:0.6)$)--++(-135:0.5) ;
	\draw ($(q1)+(135:1.2)$) node[left=-2pt]{\scriptsize $i$};
	\draw ($(q1)+(45:1.2)$) node[right=-2pt]{\scriptsize $\rho^\prime$};
	\draw ($(q1)+(-45:1.2)$) node[right=-2pt]{\scriptsize $j$};
	\draw ($(q1)+(-135:1.2)$) node[left=-2pt]{\scriptsize $\kappa$};
	
	\shade[ball color = gray, opacity=0.7] (3,-3) circle (0.6);
	\node (q2) at (3,-3) {};
	\draw[thick,decorate,decoration=snake, segment length=8pt, segment amplitude=1.5pt] ($(q2)+(45:0.6)$)--++(45:0.5) ;
	\draw[thick,decorate,decoration=snake, segment length=8pt, segment amplitude=1.5pt] ($(q2)+(-45:0.6)$)--++(-45:0.5) ;
	\draw[thick,decorate,decoration=snake, segment length=8pt, segment amplitude=1.5pt] ($(q2)+(135:0.6)$)--++(135:0.5) ;
	\draw[thick,decorate,decoration=snake, segment length=8pt, segment amplitude=1.5pt] ($(q2)+(-135:0.6)$)--++(-135:0.5) ;
	\draw[thick,double,decorate,decoration=snake, segment length=8pt, segment amplitude=1.5pt] ($(q2)+(0:0.6)$)--++(0:0.4) ;
	\draw ($(q2)+(0:1.1)$) node[right=-2pt]{\scriptsize $\rho$};
	\draw ($(q2)+(135:1.2)$) node[left=-2pt]{\scriptsize $i$};
	\draw ($(q2)+(-135:1.2)$) node[left=-2pt]{\scriptsize $j$};
	\draw ($(q2)+(45:1.2)$) node[right=-2pt]{\scriptsize $m-1$};
	\draw ($(q2)+(-45:1.2)$) node[right=-2pt]{\scriptsize $m+1$};
	\draw ($(q2)+(90:0.6)$) node[above=0pt]{\scriptsize $\ldots$};
	\draw ($(q2)+(270:0.6)$) node[below=0pt]{\scriptsize $\ldots$};

    \draw[thick,decorate,decoration=snake, segment length=8pt, segment amplitude=1.5pt] ($(q2)+(120:0.6)$)--++(120:0.45) ;
    \draw[thick,decorate,decoration=snake, segment length=8pt, segment amplitude=1.5pt] ($(q2)+(-120:0.6)$)--++(-120:0.45) ;
    \draw ($(q2)+(120:1.1)$) node[above=-2pt]{\scriptsize $i+1$};
    \draw ($(q2)+(-120:1.1)$) node[below=-2pt]{\scriptsize $j-1$};

    \node at (8.5,-3) {\scriptsize	$\xrightarrow[]{s_{m,n}=0}$ } ;
    \node at (9.5,-3.1) {	$0$ } ;
 
\end{tikzpicture}
	
    \caption{The splitting of an amplitude with $n$ legs (denoted by wavy lines) into a ``mixed" currents with $3$ $\phi$'s (denoted by straight lines) and a ``pure" current off-shell legs denoted by double lines); on the second line we show a further ``splitting" of a scalar current (such as in NLSM/sGal), which leads to a factorization near the zero contained in the four-point function.  
    }
    \label{fig:2facAnd3fac}
\end{figure*}
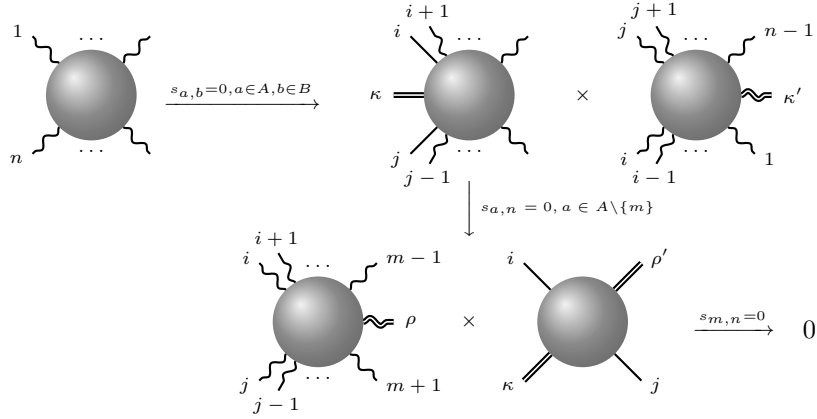

Recently, a class of hidden zeros and certain factorization behavior near such zeros have been discovered for tree-level stringy Tr $\phi^3$ amplitudes as well as deformations that give (stringy) amplitudes in non-linear sigma model and scaffolded Yang-Mills theory~\cite{Arkani-Hamed:2023swr} (see also~\cite{Arkani-Hamed:2023jry, Arkani-Hamed:2024nhp, Arkani-Hamed:2024yvu}). This has motivated the current authors to consider an even more basic type of factorization, or ``splitting" behavior, which appears to hold universally for a large class of particle and string amplitudes at tree level~\cite{Cao:2024gln}. Furthermore, in~\cite{Arkani-Hamed:2024fyd} such splitting behavior has been extended to loop level for stringy Tr $\phi^3$ and deformations in the reformulation based on ``surfaceology"~\cite{Arkani-Hamed:2023lbd, Arkani-Hamed:2023mvg}. 
Recall that one of the most familiar properties of tree-level amplitudes is the usual factorization on physical poles, where the residue of an amplitude factorizes into the product of two amplitudes with an on-shell particle (or excitations of strings) exchanged. On the other hand, this new type of factorization, or splitting, is quite different: there is no residues taken on any pole, but rather the amplitude simply factorizes on certain special loci in the kinematic space, into the product of two {\it currents} with an off-shell leg for each of them. For $n$-point amplitude, the usual factorization gives two lower-point amplitudes with $n_L + n_R=n+2$, but for splitting we have $n_L+n_R=n+3$ (see Fig~\ref{fig:2facAnd3fac}). It is quite remarkable that scattering amplitudes in a wide range of theories for colored/uncolored scalars, gauge bosons and gravitons (including their string completions) all exhibit such interesting ``splitting" behavior.

Note that this ``2-split" behavior proposed in~\cite{Cao:2024gln} and further studied here does not seem to follow from the conventional formulation of QFT scattering amplitudes based on Feynman diagrams and Lagrangians.  Remarkably, such 2-split also implies and extends more splitting/factorizing behavior studied very recently. One of them is the so-called ``smooth splitting''~\cite{Cachazo:2021wsz} (see also~\cite{GimenezUmbert:2024jjn} for resent study of $\phi^p$ theory), where certain scalar amplitudes split into three {\it currents}, when some Mandelstam variables vanish; another is the factorization near zeros proposed in~\cite{Arkani-Hamed:2023swr} which states that under certain conditions color-ordered stringy amplitudes of Tr $\phi^3$, the non-linear sigma model (NLSM) and Yang-Mills-scalar theory (YMS) all factorize into three pieces including a four-point amplitude, which in turn explains their hidden {\it zeros}; such zeros were also observed for dual resonant amplitudes in the early days of string theory~\cite{DAdda:1971wcy} and have received more attention recently~\cite{Bartsch:2024amu, Li:2024qfp}. We emphasize that all these highly non-trivial behaviors of tree-level amplitudes become simple consequences of our result: the $2$-split provides a common origin for the smooth splitting/$3$-split of~\cite{Cachazo:2021wsz} and the factorization near zeros of~\cite{Arkani-Hamed:2023swr}. Clearly the $3$-split follows from $2$-split if we apply the latter to one of the two currents again, and as shown in Fig~\ref{fig:2facAnd3fac} the factorization near zero also follows by setting some additional Mandelstam variables to zero. Furthermore, the universal applicability of our $2$-split directly generalizes all these behaviors to a much wider context~\cite{Cao:2024gln}: for example, the $3$-split is now generalized to string amplitudes and factorizations and zeros to theories without color such as the special Galileon {\it etc.}~\cite{Cachazo:2014xea}. Perhaps most importantly, by setting certain Lorentz products involving polarizations to zero (similar to Mandelstam variables), the $2$-splitting and all its consequences now directly apply to Yang-Mills and gravity amplitudes, as well as bosonic and supersymmetric string amplitudes.

As we have outlined in~\cite{Cao:2024gln}, the kinematic loci for such splittings correspond to setting a collection of Mandelstam variables to zero, and the key for such a universal behavior lies at the splitting of the universal Koba-Nielsen factor for string amplitudes (the exponential of the so-called ``scattering potentials")~\cite{Koba:1969rw}, and its saddle-point, scattering equations~\cite{Cachazo:2013gna} for particle amplitudes expressed as CHY formulas~\cite{Cachazo:2013hca, Cachazo:2013iea}. In this paper, we continue our investigations by systematically studying such splitting behavior of various string correlators and CHY integrands. For amplitudes with gluons and gravitons, similar conditions by setting Lorentz products involving polarizations are imposed as well. In this way, we will provide a rigorous proof for the splitting behavior in a web of scattering amplitudes including Z/J string integrals~\cite{Brown:2009qja,Mafra:2011nw,Mafra:2016mcc, Carrasco:2016ldy,Schlotterer:2018zce}, superstring and bosonic (open- and closed-) string amplitudes, as well as field-theory amplitudes expressed in CHY formulas for bi-adjoint $\phi^3$, non-linear sigma model (NLSM), the special Galileon (sGal), Yang-Mills scalar (YMS), Einstein-Maxwell-scalar (EMS) and Dirac-Born-Infeld (DBI), and finally also Yang-Mills and gravity amplitudes. Our results here will provide proof of this novel property for a large class of tree amplitudes from the perspective of string theory and CHY formulas.

Throughout the paper, we will use the convention that ${\cal A}_n$ denotes $n$-point particle amplitude and ${\cal M}_n$ denotes $n$-point string amplitude, which takes the general forms as follows. In the CHY formalism, any particle scattering amplitude we consider exhibits an explicit double copy structure~\cite{Bern:2008qj, Bern:2010ue, Bern:2019prr}, {\it i.e.} the integrand can be written as a product of two ``half integrands'' $I_n^\mathrm{L}/I_n^\mathrm{R}$:
\begin{align}
    \mathcal{A}_n = \int_{\mathbb{C}^n}  \dd \mu_{n}^{\mathrm{CHY}} I_n^\mathrm{L} I_n^\mathrm{R}.
\end{align}
where the CHY measure is $d\mu_n^{\rm CHY}= \prod_{a\neq i,j,k} d z_a  \delta(\partial {\cal S}_n/\partial z_a)$, and the scattering potential ${\cal S}_n$ is defined in~\eqref{eq: KN factor}. The choice of $i,j,k$ is arbitrary due to the $SL(2,\mathbb{C})$ symmetry. Without loss of generality, throughout this paper, unless specified otherwise, we choose $i < j$ and $k = n$. Similarly, the open- and closed-string integrals can be written as (with $D(\alpha)$ the open-string integration domain for ordering $\alpha$):
\begin{gather}    \mathcal{M}^{\mathrm{open}}_n=\int_{D(\alpha)} \dd \mu_n^{\mathbb{R}}\, \mathcal{I}_{n}, \\
    \mathcal{M}^{\mathrm{close}}_n=\int_{\mathbb{C}^n} \dd \mu_n^{\mathbb{C}}\, \mathcal{I}_{n}(z)\mathcal{\tilde I}_{n}(\bar z),
\end{gather}
where the open-string measure $d\mu_n^{\mathbb{R}}:= (\alpha')^{n{-}3} \prod_{a \neq i,j,k} d z_a \exp(\alpha' {\cal S}_n)$ and similarly for closed-string amplitude $d\mu_n^{\mathbb{C}}=d\mu_n^{\mathbb{R}}(z) d\mu_n^{\mathbb{R}}(\bar{z})$. 
More details about these formulas, and all the CHY (half-) integrands $I_n^\mathrm{L}/I_n^\mathrm{R}$ and string correlators $\mathcal{I}_n$ we consider in this paper will be given in section~\ref{sec:string and chy int}.

Our paper is organized as follows. In the rest of the Introduction, we review the $2$-split kinematics and show the splitting of integration measure of tree-level (open- and closed-) string amplitudes and the CHY formulas splits. In sec.~\ref{sec:string and chy int}, we give a detailed study of splitting behavior for all necessary string and CHY integrands, including the Parke-Taylor (PT) factor, CHY integrands from matrix ${\bf A}$ for Goldstone particles, and those for gluons and gravitons. In sec.~\ref{sec:stringsplit} and sec.~\ref{sec_splitParticle}, we apply these results to show how string and particle scattering amplitudes split under these conditions. Finally in sec.~\ref{sec:implications} we discuss various implications of our splitting behavior. 
\subsection{Splittings of the scattering potential and equations}

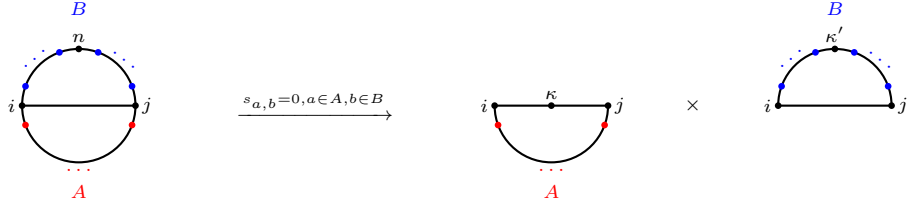
\begin{figure*}[!htbp]
    \centering
\begin{tikzpicture}[scale=1.25]
	\draw[thick] (0,0) circle (0.6);
	\node (p) at (0,0) {};
	\draw (180:0.6) node[left=-1pt]{\scriptsize $i$};
	\draw (0:0.6) node[right=-1pt]{\scriptsize $j$};
	\draw (90:0.6) node[above=-1pt]{\scriptsize $n$};
	\draw[thick] (0:0.6)--(180:0.6);
	\fill (0:0.6) circle (1pt);	\fill (180:0.6) circle (1pt);
	\fill (90:0.6) circle (1pt);

	\fill[blue] (110:0.6) circle (1pt);	
	\fill[blue] (160:0.6) circle (1pt);
	
	\fill[blue] (70:0.6) circle (1pt);	
	\fill[blue] (20:0.6) circle (1pt);
	\draw (45:0.5) node {\scriptsize \begin{rotate}{4}$\color{blue}  \ddots $\end{rotate}};
	\draw (90:0.75) node[above=4pt]{\scriptsize $\color{blue} B$};
	\draw (135:0.5) node {\scriptsize \begin{rotate}{85}$\color{blue}  \ddots $\end{rotate}};
	
	\fill[red] (-20:0.6) circle (1pt);	
	\fill[red] (200:0.6) circle (1pt);	
	\draw (-90:0.6) node[below=5pt]{\scriptsize $\color{red} A$};
	\draw (-90:0.6) node[below=-1pt]{\scriptsize $\color{red} \ldots$};

	\node at (2.5,0) {\scriptsize	$\xrightarrow[]{s_{a,b}=0,a\in A,b\in B}$};

	\node (p1) at (5,0) {};
	
	\draw[thick] ($(p1)+(-0.6,0)$) arc (-180:0:0.6);
	\draw[thick] ($(p1)+(0:0.6)$)--($(p1)+(180:0.6)$);

	\draw ($(p1)$) node[above=-1pt]{\scriptsize $\kappa$};
	\draw ($(p1)+(180:0.6)$) node[left=-1pt]{\scriptsize $i$};
	\draw ($(p1)+(0:0.6)$) node[right=-1pt]{\scriptsize $j$};

	\fill[red] ($(p1)+(-20:0.6)$) circle (1pt);	
	\fill[red] ($(p1)+(200:0.6)$) circle (1pt);	
	\draw ($(p1)+(-90:0.6)$) node[below=5pt]{\scriptsize $\color{red} A$};
	\draw ($(p1)+(-90:0.6)$) node[below=-1pt]{\scriptsize $\color{red} \ldots$};
	
	\fill ($(p1)+(0:0.6)$) circle (1pt);	\fill ($(p1)+(180:0.6)$) circle (1pt);
	\fill ($(p1)$) circle (1pt);

	\node at (6.5,0) {\scriptsize	$\times$};
	
	\node (p2) at (8,0) {};
	
	\draw[thick] ($(p2)+(-0.6,0)$) arc (180:0:0.6);
	\draw[thick] ($(p2)+(0:0.6)$)--($(p2)+(180:0.6)$);
	
	\draw ($(p2)+(90:0.6)$) node[above=-1pt]{\scriptsize $\kappa'$};
	\draw ($(p2)+(180:0.6)$) node[left=-1pt]{\scriptsize $i$};
	\draw ($(p2)+(0:0.6)$) node[right=-1pt]{\scriptsize $j$};

	\fill ($(p2)+(0:0.6)$) circle (1pt);	\fill ($(p2)+(180:0.6)$) circle (1pt);
	\fill ($(p2)+(90:0.6)$) circle (1pt);
	
	\fill[blue] ($(p2)+(110:0.6)$) circle (1pt);	
	\fill[blue] ($(p2)+(160:0.6)$) circle (1pt);
	
	\fill[blue] ($(p2)+(70:0.6)$) circle (1pt);	
	\fill[blue] ($(p2)+(20:0.6)$) circle (1pt);

	\draw ($(p2)+(45:0.5)$) node {\scriptsize \begin{rotate}{4}$\color{blue}  \ddots $\end{rotate}};
	\draw ($(p2)+(90:0.75)$) node[above=4pt]{\scriptsize $\color{blue} B$};
	\draw ($(p2)+(135:0.5)$) node {\scriptsize \begin{rotate}{85}$\color{blue}  \ddots $\end{rotate}};
\end{tikzpicture}
    \caption{$2$-split of a disk: red and blue dots denote particles in $A$ and $B$ respectively (with $A\cup B=\{1,2,\ldots,n\}\backslash\{i,j,n\}$); by imposing $s_{a\in A, b\in B}=0$ the $n$-particle system ``splits" into $\{i, j\} \cup A$ (and an off-shell $\kappa$), and $\{i,j\}\cup B$ (and an off-shell $\kappa'$). For open string or ordered particle amplitudes, the labels in two subsets preserve the ordering with $\kappa,\kappa'$ inserted at the position of $k$; Similar splitting applies to closed string and unordered amplitudes where the ordering on the disk is irrelevant. }
    \label{fig:2fac}
\end{figure*}
We define {\it 2-split} kinematics as follows: pick $3$ particles, $i,j, k$, and split the remaining $n{-}3$ into two non-empty sets $A, B$, {\it i.e.} $A\cup B=\{1,\ldots, n\}\backslash\{i,j,k\}$, then we demand 
\begin{align}
 \label{eq_splitKin}
    s_{a,b} = 0,\qquad \forall a \in A, b \in B,
\end{align} 
where the Mandelstam invariants is $s_{a,b,\ldots,d}= (p_a + p_b + \ldots + p_d)^2$. We want to show that under this condition, the amplitudes in various scalar theories split into two amputated currents, each with one off-shell leg. 

We start with the scattering potential, or $\log$ of the Koba-Nielsen factor:
\begin{equation} \label{eq: KN factor}
{\cal S}_n=\sum_{a<b} s_{a,b} \log z_{a,b}=\sum_{a<b\neq k, (a,b) \neq (i,j)} s_{a,b} \log |ab|   
\end{equation}
where we have defined $z_{a,b}:=z_b-z_a$; in the second equality, we have rewritten the potential by solving $s_{a,k}$ for $a\neq k$ as well as $s_{i,j}$ in terms of the remaining $n(n{-}3)/2$ independent $s_{a,b}$, and we have also defined the SL$(2)$ invariant: $|ab|:=\frac{z_{a, b} z_{i, k} z_{j, k}}{z_{a, k} z_{b, k} z_{i, j}}$. Furthermore, we can fix the  SL$(2)$ redundancy by setting $z_i=0, z_j=1$ and $z_k \to \infty$, such that the $\mathrm{SL}(2)$ invariant is simplified as $|ab|=z_{a,b}$. Under the 2-split kinematics~\eqref{eq_splitKin}, it is straightforward to see that \eqref{eq: KN factor} naturally splits into ``left" and ``right" parts:
\begin{equation}
{\cal S}_n \to ({\cal S}_A + {\cal S}_{i, A} + {\cal S}_{j, A}) + ({\cal S}_B + {\cal S}_{i, B} + {\cal S}_{j, B}):={\cal S}_L(i,j,A; \kappa)+ {\cal S}_R(i,j,B; \kappa'),    
\end{equation}
where we have ${\cal S}_A=\sum_{a<b, a,b \in A} s_{a,b} \log |ab|$, ${\cal S}_{i, A}=\sum_{a\in A} s_{i,a} \log |i a|$, ${\cal S}_{j, A}=\sum_{a\in A} s_{a, j} \log |a j|$ (in the chosen gauge fixing, $|ia|=z_a-z_i=z_a$, $|aj|=z_j-z_a=1-z_a$) and similarly for the right part. As shown in the second equality of the above formula, we can package the terms involving set $A$/$B$ together and re-interpret the two sums as the scattering potentials for two currents: the first one (left) with on-shell legs $a\in A$, $i, j$ and an off-shell leg $\kappa$ with momentum $p_\kappa=-\sum_{a\in A} p_a -p_i-p_j$; the second one (right) with on-shell legs $b\in B, i, j$ and an off-shell leg $\kappa'$ with momentum $p_{\kappa'}=-\sum_{b\in B} p_b-p_i-p_j$ (see Figure~\ref{fig:2fac}). In this way, momentum conservation is preserved separately for each of the current, and the left/right one contains $|A|+3$ and $|B|+3$ external legs respectively, so in total we have $n+3$ legs; the dimensions of these moduli spaces thus add up: $n{-}3=|A|+|B|=n_L{-}3+n_R{-}3$. We note that, in principle, $z_i, z_j$ for the left and right currents should be taken as independent variables since they live in the respective moduli spaces, but since we have fixed $z_i=0, z_j=1$ and both $z_\kappa, z'_{\kappa} \to \infty$, this can be neglected. Our gauge-fixing choice also breaks the symmetry between $i,j$ and $k$, and we have chosen $i, j$ to be on-shell in both currents, meaning that the remaining $\kappa/\kappa'$ (that replaces leg $k$) must be off-shell; we could equally make other choices for off-shell legs. 
Consequently, the open-string/close-string/CHY measure also splits into  ``left'' and ``right'' parts
\begin{equation}
\dd \mu_n=\dd \mu_L (i,j, A; \kappa) \dd\mu_R (i,j, B; \kappa')\,,\quad {\rm for}~\dd\mu_n^{\mathbb{R}}, \dd\mu_n^{\mathbb{C}}~{\rm and}~\dd\mu_n^{\rm CHY}\,.
\end{equation}

In section~\ref{sec:string and chy int}, we show that the CHY/string integrands and the integration domains for open-string amplitudes also split in the 2-split kinematics~\eqref{eq_splitKin}, and for the gauge/gravity theory, we need more conditions for the Lorentz products ($\epsilon \cdot \epsilon,\epsilon \cdot p$) to split the integrand with the spinning particles. We define the extra conditions for 2-split of gluon,
\begin{equation}\label{eq_pol}
	\epsilon_a \cdot \epsilon_{b'}=0\,, \quad p_a \cdot \epsilon_{b'}=0\,,\quad \epsilon_a \cdot p_b=0\,,
\end{equation}
where $a \in A, b \in B$, and $b^{\prime} \in B^{\prime}=B\cup \{i,j,k\}$. For graviton, the conditions are similar for $\epsilon$ and $\tilde{\epsilon}$.

Let us summarize the main results here, with detailed discussions provided in the following sections. In general, the scalar amplitudes from CHY formalism or string integral split into two currents under the 2-split kinematics, depicted in Fig~\ref{fig:2facAnd3fac}, 
\begin{equation}
\mathcal{A}^{\mathrm{scalar}}_n/ \mathcal{M}^{\mathrm{open}}_n/ \mathcal{M}^{\mathrm{close}}_n\xrightarrow[]{s_{a,b}=0,a\in A,b\in B}\mathcal{J}^{\mathrm{mixed}}_{j-i+2}(i^\phi, A, j^\phi; \kappa^\phi)\times\mathcal{J}_{n-j+i+1}(j,B,i;\kappa'),
\end{equation}
where $\mathcal{J}_n(\ldots; \kappa)$ denotes the particle/string current with off-shell leg $\kappa$, and the superscript ``mixed'' means that $\{i,j, \kappa\}$ are different types of particle/string state (\textit{e.g.} $\phi^3$) than the original ones.
For color-ordered amplitudes, the results of the splitting are the product of ordered currents with an off-shell leg which takes the position of that of the leg $k=n$, we use the semicolon to single out the off-shell leg, which is not written according to the ordering\footnote{The expressions for ordered currents that respect the ordering with $k=n$ read $\mathcal{J}^{\mathrm{mixed}}_{j-i+2}(i^\phi, A, j^\phi, \kappa^\phi)$ and $\mathcal{J}_{n-j+i+1}(j,\ldots,n-1,\kappa',1,\ldots,i)$}.

The gluon amplitudes from CHY formalism or open-string integral split into two currents under the 2-split kinematics and the extra conditions defined in~\eqref{eq_pol},
\begin{equation}
\mathcal{A}^{\mathrm{YM}}_n/ \mathcal{M}^{\mathrm{open}}_n\xrightarrow[\eqref{eq_pol}]{s_{a,b}=0,a\in A,b\in B}{\cal J}^{\mathrm{YM}+\phi^3}_{j-i+2} (i^\phi, A, j^\phi; \kappa^\phi) \times {\cal J}^{\mathrm{YM}}_{n-j+i+1,\mu} (j, B, i; \kappa') \epsilon_n^\mu\,,  
\end{equation}
where the ``pure" gluon current has an off-shell leg $\kappa'$, contracted with polarization $\epsilon_n$.

The gravity amplitudes from CHY formalism or close-string integral have two different 2-split behaviors depending on the choices of extra conditions~\eqref{eq_pol} for $\epsilon$ and $\tilde{\epsilon}$.
\begin{equation}\label{eq_GR split1}
\mathcal{A}^{\mathrm{GR}}_n/ {\cal M}^{\rm closed}_n \xrightarrow[\tilde{\epsilon}_a \cdot \tilde{\epsilon}_{b'},p_a \cdot \tilde{\epsilon}_{b'},\tilde{\epsilon}_a \cdot p_b=0]{p_a\cdot p_b,\epsilon_a \cdot \epsilon_{b'},p_a \cdot \epsilon_{b'},\epsilon_a \cdot p_b=0} {\cal J}^{\mathrm{GR}+\phi^3}_{j-i+2} (i^\phi, A, j^\phi; \kappa^\phi) \times {\cal J}^\mathrm{GR}_{n-j+i+1,\mu\nu} (j, B, i; \kappa') \epsilon_n^\mu \tilde{\epsilon}_n^\nu\,. 
\end{equation}
\begin{equation}\label{eq_GR split2}
\mathcal{A}^{\mathrm{GR}}_n/ {\cal M}^{\rm closed}_n \xrightarrow[\tilde{\epsilon}_{a'} \cdot \tilde{\epsilon}_{b},p_a \cdot \tilde{\epsilon}_{b},\tilde{\epsilon}_{a'} \cdot p_b=0]{p_a\cdot p_b,\epsilon_a \cdot \epsilon_{b'},p_a \cdot \epsilon_{b'},\epsilon_a \cdot p_b=0} {\cal J}^{\rm EYM}_{j-i+2,\nu} (i^g, A, j^g; \kappa^g) \tilde{\epsilon}_n^\nu \times {\cal J}^\mathrm{EYM}_{n-j+i+1,\mu} (j^g, B, i^g; \kappa^{g,\prime}) \epsilon_n^\mu \, ,
\end{equation}
where  $a' \in A \cup \{i,j,k\}$. The EYM denotes Einstein-Yang-Mills theory~\cite{Cachazo:2014nsa,Cachazo:2014xea}. The $A,B$ are gravitons, and $(i^g,j^g,\kappa^g,\kappa^{g,\prime})$ are gluons.

\section{Splittings of string correlators and CHY (half) integrands}\label{sec:string and chy int}
In this section, we provide a comprehensive derivation of splitting behavior for all necessary ingredients for splittings of a large class of tree-level amplitudes represented by string/CHY integrals.
Let us begin by defining the string/particle amplitudes. The string(y) amplitudes of interest include:
\begin{itemize}
    \item \emph{Stringy Bi-adjoint $\phi^3$}: There are two $\alpha'$ completions of the particle bi-adjoint $\phi^3$ theory, formulated in open- and closed-string fashions~\cite{Brown:2009qja,Mafra:2011nw,Mafra:2016mcc,Schlotterer:2018zce}:
    \begin{equation}
        Z_{\alpha|\beta} := \int_{D(\alpha)} \dd \mu_n^{\mathbb{R}}\, \mathrm{PT}(\beta)\,, \quad J_{\alpha|\beta} := \int_{\mathbb{C}^n} \dd \mu_n^{\mathbb{C}}\, \mathrm{PT}(\alpha)\mathrm{PT}(\beta)\,,
    \end{equation}
    referred to as the $Z$ and $J$ integrals, respectively. Here, $\mathrm{PT}(\alpha)$ is the Parke-Taylor factor, which is defined in \eqref{eq_PT}.
    
    \item \emph{Bosonic String Amplitudes}~\cite{Green:1987sp}: The tree-level open- and closed-string amplitudes are given by:
    \begin{gather}
        \mathcal{M}_n^{\mathrm{bosonic,open}} = \int_{D(\alpha)} \dd \mu_n^{\mathbb{R}}\, \mathcal{C}_n (\{\epsilon, p, z\}), \\
        \mathcal{M}_n^{\mathrm{bosonic,close}} = \int_{\mathbb{C}^n} \dd \mu_n^{\mathbb{C}}\, \mathcal{C}_n (\{\epsilon, p, z\})\, \mathcal{C}_n (\{\tilde{\epsilon}, p, \bar{z}\}),
    \end{gather}
    where the bosonic string correlator $\mathcal{C}_n (\{\epsilon, p, z\})$ is defined in equation \eqref{eq_bosonic correlator}.
    
    \item \emph{Superstring Amplitudes}~\cite{Green:1987sp}: The tree-level open superstring amplitude is:
    \begin{gather}
        \mathcal{M}_n^{\mathrm{type-I}} = \int_{D(\alpha)} \dd \mu_n^{\mathbb{R}}\, \varphi^{\mathrm{type-I}}_{n} (\{\epsilon, p, z\}), \\
        \mathcal{M}_n^{\mathrm{type-II}} = \int_{\mathbb{C}^n} \dd \mu_n^{\mathbb{C}}\, \varphi^{\mathrm{type-I}}_{n}(\{\epsilon, p, z\})\varphi^{\mathrm{type-I}}_{n} (\{\tilde\epsilon, p, \bar z\}),
    \end{gather}
    where the superstring correlator $\varphi^{\mathrm{type-I}}_{n} (\{\epsilon, p, z\})$ is defined in equation \eqref{eq_superstring correlator}.
\end{itemize}
Here, $\alpha$ and $\beta$ represent permutations of $(1, 2, \ldots, n)$, and the open-string integration domain is $\mathcal{D}(\alpha):=\left\{\left(z_1, z_2, \ldots, z_n\right) \in \mathbb{R}^n \mid-\infty<z_{\alpha_1}<z_{\alpha_2}<\ldots<z_{\alpha_n}<\infty\right\}$.

For particle amplitudes, we study theories that admit a CHY representation, which allows us to write tree-level amplitudes for massless particles as
\begin{align}
   \mathcal{A}_n = \int_{\mathbb{C}^n}  \dd \mu_n^{\mathrm{CHY}}I_n^\mathrm{L} I_n^\mathrm{R},
\end{align}
where the theory-dependent $I^{L}_{n}, I^{R}_{n}$ are referred to as half integrands. In this paper, we consider four distinct half integrands: (1) the Parke-Taylor factor $\mathrm{PT}(\alpha)$ defined in \eqref{eq_PT}; (2) the reduced determinant $\mathrm{det}^\prime \mathbf{A}_n$ \eqref{eq_detA}; (3) the reduced Pfaffian  $\mathrm{Pf}^\prime \mathbf{\Psi}_n$ \eqref{eq_PfPsi}; (4) $\mathrm{Pf}^\prime \mathbf{A} _n\mathrm{Pf} \mathbf{X}_n$ defined in~\eqref{eq:pfA} and~\eqref{eq: PfX as sum}. These are typically functions of the external momenta, polarizations, and orderings. To avoid distraction, we delay their definitions to the subsections where we discuss their splitting behaviors. For now, we focus on the CHY integrands that can be constructed from these four half integrands.
In the spirit of the double copy, we collect the ten theories in Table \ref{tab_CHY}.
\begin{table}[!h]
    \begin{center}
        \caption{Integrands for CHY formula.}
        \label{tab_CHY}
        \begin{tabular}{|c|c|c|c|c|}
            \hline
            &
            $\mathrm{PT}(\alpha)$ &
            $\mathrm{det}^\prime \mathbf{A}_n$ &
            $\mathrm{Pf}^\prime \mathbf{\Psi}_n$ &
            $\mathrm{Pf}^\prime \mathbf{A}_n \mathrm{Pf}  \mathbf{X}_n$ \\
            \hline
            $\mathrm{PT}(\beta)$ & Bi-adjoint $\phi^3$ & NLSM & YM & YMS \\
            \hline
            $\mathrm{det}^\prime \mathbf{A}_n$ & NLSM & sGal & BI & DBI \\
            \hline
            $\mathrm{Pf}^\prime \mathbf{\Psi}_n$ & YM & BI & GR & EM \\
            \hline
            $\mathrm{Pf}^\prime \mathbf{A}_n \mathrm{Pf}  \mathbf{X}_n$ & YMS & DBI & EM & EMS \\
            \hline
        \end{tabular}
    \end{center}
\end{table}
To be concrete, we explicitly spell out the amplitudes as (most of the original definitions can be found in \cite{Cachazo:2013iea,Cachazo:2014xea})
\begin{itemize}
    \item \emph{Particle Bi-adjoint $\phi^3$}: 
        $\mathcal{A}_n^{\phi^3} = \int \dd \mu_n^{\mathrm{CHY}} \mathrm{PT}(\alpha) \, \mathrm{PT}(\beta)$. \\
        It can be recovered from the Z/J integral by taking the field-theory limit $\alpha^\prime \to 0$. 
    \item \emph{Non-linear $\sigma$ model (NLSM)}: 
        $\mathcal{A}_n^{\mathrm{NLSM}} = \int \dd \mu_n^{\mathrm{CHY}} \mathrm{PT}(\alpha) \, \mathrm{det}^\prime \mathbf{A}_n$. 
        \\
        It has the $\mathrm{U}(N)$ flavour group ({\it c.f.}~\cite{Kampf:2013vha}).
    \item \emph{Yang-Mills (YM)}: 
        $\mathcal{A}_n^{\mathrm{YM}} = \int \dd \mu_n^{\mathrm{CHY}} \mathrm{PT}(\alpha) \, \mathrm{Pf}^\prime \mathbf{\Psi}_n$.
    \item \emph{Yang-Mills-scalar (YMS)}: 
        $\mathcal{A}_n^{\mathrm{YMS}} = \int \dd \mu_n^{\mathrm{CHY}} \mathrm{PT}(\alpha) \, \mathrm{Pf}^\prime \mathbf{A}_n \mathrm{Pf}  \mathbf{X}_n$. \\
        This is a theory of bi-adjoint scalars coupled to gluons. It can be derived from the dimensional reduction of pure YM theory.
    \item \emph{Special Galileon (sGal)}: 
        $\mathcal{A}_n^{\mathrm{sGal}} = \int \dd \mu_n^{\mathrm{CHY}} \mathrm{det}^\prime \mathbf{A}_n \,  \mathrm{det}^\prime \mathbf{A}_n$. 
    \item \emph{Born-Infeld (BI)}: 
        $\mathcal{A}_n^{\mathrm{BI}} = \int \dd \mu_n^{\mathrm{CHY}} \mathrm{det}^\prime \mathbf{A}_n \, \mathrm{Pf}^\prime \mathbf{\Psi}_n$.
    \item \emph{Dirac-Born-Infeld (DBI)}: 
        $\mathcal{A}_n^{\mathrm{DBI}} = \int \dd \mu_n^{\mathrm{CHY}} \mathrm{det}^\prime \mathbf{A}_n \, \mathrm{Pf}^\prime \mathbf{A}_n \mathrm{Pf} \mathbf{X}_n$.
    \item \emph{Gravity (GR)}: 
        $\mathcal{A}_n^{\mathrm{GR}} = \int \dd \mu_n^{\mathrm{CHY}} \mathrm{Pf}^\prime \mathbf{\Psi}_n \, \mathrm{Pf}^\prime \mathbf{\Psi}_n$.
    \item \emph{Einstein-Maxwell (EM)}: 
        $\mathcal{A}_n^{\mathrm{EM}} = \int \dd \mu_n^{\mathrm{CHY}} \mathrm{Pf}^\prime \mathbf{\Psi}_n \, \mathrm{Pf}^\prime \mathbf{A}_n \mathrm{Pf} \mathbf{X}_n$.
    \item \emph{Einstein-Maxwell-scalar (EMS)}: 
        $\mathcal{A}_n^{\mathrm{EMS}} = \int \dd \mu_n^{\mathrm{CHY}} \mathrm{Pf}^\prime \mathbf{A}_n \mathrm{Pf} \mathbf{X}_n \, \mathrm{Pf}^\prime \mathbf{A}_n \mathrm{Pf} \mathbf{X}_n$.
\end{itemize}
In addition to these pure amplitudes, we will encounter some mixed amplitudes that arise from splittings of pure amplitudes~\cite{Cachazo:2016njl}. For example, as we will see, a pure pion amplitude will split into a lower-point pure pion current and a mixed one involving three bi-adjoint scalars and the remaining particles being pions. All these mixed amplitudes can be defined using the CHY formalism. 
We are now ready to examine the splitting behavior of the string correlators and CHY half-integrands.

\subsection{Parke-Taylor factor}\label{sec:PT split}
Firstly, let us consider the Parke-Taylor factor that appears in both stringy and particle amplitudes.
Given a specific color ordering $\alpha \in \mathbf{S}_n$, we define
\begin{align}
    \label{eq_PT}
    \mathrm{PT}(\alpha) = \frac{\Delta_{i,j,k}} {z_{\alpha_1,\alpha_2} z_{\alpha_2,\alpha_3} \ldots z_{\alpha_{n{-}1},\alpha_n} z_{\alpha_n,\alpha_1}}, 
\end{align}
where we have absorbed the Jacobian 
\begin{align}
    \Delta_{i,j,k} = z_{i,j} z_{j,k} z_{k,i}.
\end{align}
from gauge-fixing into the Parke-Taylor factor.
For simplicity, we focus on the canonical ordering $\mathbb{I} = (1, 2, \ldots, n)$ and choose $ i<j<k=n$; other orderings can be obtained through relabeling\footnote{The ordering must be ``compatible'' with the split kinematics such that the PT factor will split in the same way as the measure, {\it i.e.}, the elements in the set $A$ cannot be adjacent to those in the set $B$.}. The splitting of the Parke-Taylor factor has been comprehensively discussed in~\cite{Cao:2024gln} (see also~\cite{Cachazo:2021wsz}).
Here, we provide a concise recap for completeness. It is evident that once the $SU(2)$ gauge is fixed to $z_i = 0, z_j = 1, z_k = z_n = \infty$, the Parke-Taylor factor simplifies to
\begin{align}
    \mathrm{PT}(I) = \frac{1}{z_{{i},{i{+}1}} \ldots z_{{j{-}1},j}} \times
    \frac{1}{z_{j,{j{+}1}} \ldots z_{{n{-}2},{n{-}1}} \times z_{1,2} \ldots  z_{{i{-}1},i}} \, ,
\end{align}
The key observation is that if we insert $z_{\kappa} = \infty$ and $z_{\kappa^\prime} = \infty$ and restore $z_i = 0, z_j = 1$, the left and right factors become\footnote{Strictly speaking, $z_i$/$z_j$ for the left and right factors should be taken as independent punctures if one wishes to restore full $SU(2)$ invariance.}
\begin{gather}
    \mathrm{PT} (i, i{+}1, \ldots, j{-}1, j, \kappa) = \frac{\Delta_{i,j,\kappa}}{z_{{i},{i{+}1}} \ldots z_{{j{-}1},j} z_{j,\kappa} z_{\kappa,i}}. \\
    \mathrm{PT} (j, j{+}1, \ldots, n{-}1, \kappa^\prime, 1, \ldots,i{-}1, i) = 
    \frac{\Delta_{i,j,\kappa^\prime}}{z_{j,{j{+}1}} \ldots z_{{n{-}2},{n{-}1}} z_{{n{-}1}, \kappa^\prime} z_{\kappa^\prime, 1}  z_{1,2} \ldots  z_{{i{-}1},i}  z_{i,j}}\,.
\end{gather}
Therefore, the $n$-point PT factor splits into two parts, each being another PT factor for lower points:
\begin{align} \label{eq: splitting of PT}
    \mathrm{PT}(I) = \mathrm{PT}(i,(i,j),j,\kappa) \times \mathrm{PT}(j, (j,n), \kappa', (n,i), i)\,,
\end{align}
where, for brevity, we have used $(i,j)$ to denote the ordered set $\{i+1, \ldots, j-1\}$, and similarly for $(j,n)$ and $(n,i)$ (cyclically).
This factorization separates the particles (excluding $\{i, j, k\}$) into two sets, with $\{i, j\}$ forming the border between them. However, this splitting is mostly artificial. Alternatively, we can choose $\{i,k\}$ or $\{j,k\}$ as the boundary, and the Parke-Taylor factor will factorize accordingly. The Parke-Taylor factor can also be further decomposed into three parts, leading to the three splits as demonstrated in~\cite{Cachazo:2021wsz}. Finally, we note that this splitting is independent of the kinematics and thus completely universal.

\subsection{CHY integrands from matrix $\mathbf{A}$ for Goldstone particles} \label{sec: splitting of matrix A}
As we have mentioned, a crucial ingredient in CHY formulas for NLSM, DBI and sGal (known as ``exceptional EFT"~\cite{Cheung:2014dqa}) which encodes the enhanced Adler's zero~\cite{Adler:1964um,Cheung:2014dqa,Cheung:2018oki} is the reduced determinant or Pfaffian of the matrix $\mathbf{A}_n$.  This is an $n \times n$ anti-symmetric matrix with entries
\begin{align}
    (\mathbf{A}_n)_{a b}=\begin{cases}
        \frac{2p_a \cdot p_b}{z_a-z_b} & a \neq b, \\
        0 & a=b.
    \end{cases}
\end{align}
On the support of the scattering equations, $\mathbf{A}_n$ has co-rank $2$. 
\paragraph{${\rm det}'\mathbf{A}_n$} We define the reduced determinant as\footnote{We note that we absorb $\Delta_{i,j,k}$ into the definition, which is different from one usually finds in the literature.}
\begin{equation}
    \label{eq_detA}
    \det{}^\prime \mathbf{A}_n := \Delta_{i,j,k} \frac{(-1)^{p+q}}{(z_p - z_q)^2} \det\mathbf{A}_n^{[p,q]},
\end{equation}
where $\mathbf{A}_n^{[p,q]}$ denotes the minor with the $p^\mathrm{th}$ and $q^\mathrm{th}$ rows and columns removed. Throughout this paper, we will drop the sign $(-1)^{p+q}$ in the definition of reduced determinant and reduced Pfaffian to keep the expressions clean.  We note that $\det{}^\prime \mathbf{A}_n$ is independent of the choice of $p,q$. For simplicity, we take $p=j,\, q=n$. 
 Now let us consider the behavior of det$^\prime \mathbf{A}_n$ under the splitting kinematics~\eqref{eq_splitKin}. Recall we must have even value of $n$ otherwise the half integrand vanishes. In addition, $i,j,k=n$ have been singled out and split the rest particles into two sets $(i,j)=\{i+1,\ldots,j-1\}$ and $(j,n)\cup (n,i)=\{j+1,\ldots,n-1,1,\ldots,i-1\}$, one of which contains an even number of elements and the other has odd.
Without losing generality, we presume that $|(i,j)|$ is odd and $|(j,n)\cup (n,i)|$ is even; the other case can be treated analogously by exchanging the two sets. We can write the $(n-2) \times (n-2)$ submatrix $\mathbf{A}_n^{[j,n]}$ into the following form:
\begin{equation}
    \mathbf{A}_n^{[j,n]} \to \left(
    \begin{array}{cc|c}
        \large{\mathbf{A}}_{(i, j)}& \begin{array}{c}
            \mathbf{A}_{i+1,i}   \\
            \mathbf{A}_{i+2,i}   \\
            \vdots               \\
            \mathbf{A}_{j-1,i}   \\
        \end{array} & \large{0} \\
        \begin{array}{cccc}
            \mathbf{A}_{i,i+1} & \mathbf{A}_{i,i+2} &\ldots &\mathbf{A}_{i,j-1}
        \end{array} & 0
        & \begin{array}{cccccc} \mathbf{A}_{i,j+1} & \ldots& \mathbf{A}_{i,n-1} & \mathbf{A}_{i,1} & \ldots & \mathbf{A}_{i,i-1}
        \end{array} \\
        \hline
        \large{0} &\begin{array}{c}
            \mathbf{A}_{j+1,i}   \\
            \vdots   \\
            \mathbf{A}_{n-1,i}               \\
            \mathbf{A}_{1,i}   \\
            \vdots \\
            \mathbf{A}_{i-1,i}   \\
        \end{array}& \large{\mathbf{A}}_{(j,n)\cup(n,i)}
    \end{array}
    \right)
\end{equation}
where $\mathbf{A}_{(i, j)}$ and $\mathbf{A}_{(j,n)\cup(n,i)}$ denote the submatrices with rows and columns from their respective subscript.
By exploiting the permutation invariance of the determinant (up to a sign), we move the $i^\mathrm{th}$ row and column as the boundary between the two submatrices. 
Using a simple lemma in~\cite{Cachazo:2021wsz}, it is straightforward to see:
\begin{equation} \label{eq:A_split_odd}
    \mathrm{det} \mathbf{A}_n^{[j,n]}\to  \mathrm{det} \mathbf{A}_{\{i\} \cup (i,j)} \;  \mathrm{det} \mathbf{A}_{(j,n)\cup(n,i)}.
\end{equation}
Consequently, together with the prefactor in \eqref{eq_detA}, and restoring the $SU(2)$ invariance as we did for the PT factor, $\mathrm{det}^\prime \mathbf{A}_n$ splits as
\begin{equation}
    \mathrm{det}^\prime \mathbf{A}_n \to  \left(\Delta_{i,j,\kappa} \mathrm{PT}(j,\kappa) \mathrm{det} \mathbf{A}_{\{i\} \cup (i,j)} \right) \times \left(\Delta_{i,j,\kappa'} \mathrm{PT}(i,j,\kappa') \mathrm{det} \mathbf{A}_{(j,n)\cup(n,i)} \right).
\end{equation}
As we will see in section~\ref{sec_splitParticle}, when inserted into the CHY formula, on the right-hand side, the first factor corresponds to a pure amplitude with particles $(i,j)\cup \{i,j,\kappa\}$, whereas the second factor corresponds to a mixed amplitude with $(j,n)\cup (n,i)$ being the original particles but $(i,j,\kappa)$ of another kind (for example, bi-adjoint scalar in the splitting of a pure pion amplitude).

\paragraph{$\mathrm{Pf}^\prime \mathbf{A}_n \mathrm{Pf} \mathbf{X}_n $}
Another half integrand that involves the matrix $\mathbf{A}_n$, which is a key ingredient for the YMS/DBI/EMS amplitudes, is $\mathrm{Pf}^\prime \mathbf{A}_n \mathrm{Pf} \mathbf{X}_n $.
Then reduced Pfaffian $\mathrm{Pf}^\prime \mathbf{A}_n$ is defined by
\begin{equation}\label{eq:pfA}
    \operatorname{Pf}^{\prime} \mathbf{A}_n := \Delta_{i,j,k} \frac{(-1)^{p+q}}{\left(z_p-z_q\right)} \operatorname{Pf}\left(\mathbf{A}_n^{[p,q]}\right).
\end{equation}
Again, we can fix $p=j, q=n$.
Besides, we define $\mathbf{X}_n$ as the $n \times n$ anti-symmetric matrix:
\begin{equation}
(\mathbf{X}_n)_{a b}=\begin{cases}
    \frac{1}{z_a-z_b} & a \neq b, \\
    0 & a=b.
\end{cases} 
\end{equation}
Practically, one usually decomposes the Pfaffian of $\mathbf{X}_n$ as:
\begin{equation} \label{eq: PfX as sum}
    \operatorname{Pf} \mathbf{X}_n =  \sum_{t\, \in \text {p.m.}} \operatorname{sgn}(t) \frac{1}{z_{t_1, t_2}z_{t_3, t_4} \ldots z_{t_{n-1},t_n}},
\end{equation}
where we sum over all perfect matching (p.m.) $t$ with each term weighted by a sign $\operatorname{sgn}(t)$~\cite{Cachazo:2014xea}. Crucially, when we discuss the splitting of this object, we focus on one specific term in the sum since as we will soon explain below, the splitting kinematics need to be compatible with the perfect matching.

Let us consider one term in the decomposition~\eqref{eq: PfX as sum} denoted by $\mathrm{Pf}^\prime \mathbf{A}_n/( z_{t_1,t_2} \ldots z_{t_{n-1},t_n})$. For simplicity, we neglect the sign of this term. Firstly, note the $\mathrm{Pf}^\prime \mathbf{A}_n$ behaves exactly the same as $\mathrm{det}^\prime \mathbf{A}_n$ in the split kinematics:
\begin{equation}
    \label{eq_PFAsplit}
    \mathrm{Pf} \mathbf{A}_n^{[j,n]}\to  \mathrm{Pf} \mathbf{A}_{\{i\} \cup (i,j)} \;  \mathrm{Pf} \mathbf{A}_{(j,n)\cup(n,i)},
\end{equation}
where again we have assumed that $|(i,j)|$ is odd and $|(j,n)\cup(n,i)|$ is even. Crucially, we require that the factor $1/\left(z_{t_1,t_2} \ldots z_{t_{n-1},t_n}\right)$ does not contain any such pair $z_{t_{p},t_{p+1}}$ with one element from $(i,j)$ and the other from $(j,n)\cup(n,i)$. In other words, we need $\{t_p,t_{p+1}\} \subset (i,j)$ or $\{t_p,t_{p+1}\} \subset (j,n)\cup(n,i)$ for any pair. By this requirement, the pairs that do not involve $\{i,j,n\}$ trivially split. Which elements $\{i,j,n\}$ are paired with then characterizes the splitting of this factor. Let $a_p$ denote a label in a given permutation of $(i,j)$, and $b_q$ in a permutation of $(j,n)\cup(n,i)$, respectively.
Careful analysis shows that there are only three distinct cases: (1) two of $\{i,j,n\}$ are paired together, then the last, \textit{e.g.} $i$, must be paired with one element $a_1$ in $(i,j)$ (note that $|(i,j)|$ is odd); (2) $\{i,j,n\}$ are all paired with elements in $(i,j)$; (3) one of $\{i,j,n\}$, say $n$, is paired with an element in $(i,j)$, and $j,n$ are separately paired with elements in $(j,n)\cup(n,i)$. 
Taking into account of the prefactor $\Delta_{i,j,n}/z_{n,j}$ from \eqref{eq:pfA}, all these three cases splits
\begin{subequations}
\label{eq:YMSallcases}
\begin{align}
    (1):& \qquad \frac{\Delta_{i,j,n}}{z_{n,j}} \frac{1}{z_{i,a_1} z_{j,n}} \to \frac{1}{z_{i,a_1}} \mathrm{PT}(j,\kappa) \times \mathrm{PT}(i,j,\kappa^\prime), 
    \label{eq:YMScase1} \\
    (2):& \qquad 
    \frac{\Delta_{i,j,n}}{z_{n,j}} \frac{1}{z_{i,a_1} z_{j,a_2}z_{n,a_3}} \to \frac{\Delta_{i,j,\kappa} }{z_{i,a_1}z_{j,a_2}z_{\kappa,a_3}z_{j, \kappa}} \times \mathrm{PT}(i,j,\kappa^\prime),
    \label{eq:YMScase2} \\
    (3):& \qquad 
    \frac{\Delta_{i,j,n}}{z_{n,j}} \frac{1}{z_{i, b_1} z_{j, b_2}z_{n, a_1}} \to \frac{\Delta_{i,j,\kappa} }{z_{i, j}z_{j, \kappa}z_{\kappa, a_1}}  \times \frac{\Delta_{i,j,\kappa^\prime}}{z_{i, b_1}z_{j, b_2}z_{i, \kappa^\prime} z_{j, \kappa^\prime}}.
    \label{eq:YMScase3}
\end{align}
\end{subequations}
Finally, combining \eqref{eq_PFAsplit} and \eqref{eq:YMSallcases}, we find that a given perfect matching component of the half integrand $\mathrm{Pf}^\prime \mathbf{A}_n \mathrm{Pf} \mathbf{X}_n $ splits as

{\allowdisplaybreaks[1]
\begin{subequations}
\begin{align} \label{eq2:YMScase1}
&\begin{aligned}
    (1): \qquad &\frac{\mathrm{Pf}^\prime \mathbf{A}_n}{z_{i, a_1} z_{j,n}\prod_{p}z_{a_p, a_{p+1}} \prod_{q}z_{b_q, b_{q+1}}} \to\\
    &   \mathrm{PT}(j,\kappa) \frac{\mathrm{Pf} \mathbf{A}_{\{i\} \cup (i,j)}}{z_{i,a_1}\prod_{p}z_{a_p, a_{p+1}}}
    \times  \mathrm{PT}(i,j,\kappa^\prime) \frac{\mathrm{Pf} \mathbf{A}_{(j,n)\cup(n,i)}}{\prod_{q}z_{b_q, b_{q+1}}}\\
    =&\Delta_{i,j,\kappa}   \frac{\mathrm{Pf} \mathbf{A}^{[j,\kappa]}_{\{i,j,\kappa\} \cup (i,j)}/z_{j ,\kappa}}{z_{j, \kappa} z_{i,a_1}\prod_{p}z_{a_p, a_{p+1}}}
    \times   \mathrm{PT}(i,j,\kappa^\prime) \frac{\mathrm{Pf} \mathbf{A}_{(j,n)\cup(n,i)}}{\prod_{q}z_{b_q, b_{q+1}}}
    ,
\end{aligned} \\
&
\label{eq2:YMScase2}
\begin{aligned}
    (2): \qquad\qquad &\frac{ \mathrm{Pf}^\prime \mathbf{A}_n}{z_{i,a_1} z_{j,a_2}z_{k,a_3}\prod_{p}z_{a_p,a_{p+1}} \prod_{q}z_{b_q, b_{q+1}}} \to\\
    &\Delta_{i,j,\kappa}  \frac{\mathrm{Pf} \mathbf{A}_{\{i\} \cup (i,j)}}{z_{i,a_1}z_{j,a_2}z_{\kappa,a_3}z_{j,\kappa}\prod_{p}z_{a_p,a_{p+1}}}
    \times  \mathrm{PT}(i,j,\kappa^\prime) \frac{\mathrm{Pf} \mathbf{A}_{(j,n)\cup(n,i)}}{\prod_{q}z_{b_q, b_{q+1}}}\\
    = &\Delta_{i,j,\kappa}  \frac{\mathrm{Pf} \mathbf{A}^{[j,\kappa]}_{\{i,j,\kappa\} \cup (i,j)}/z_{j ,\kappa}}{z_{i,a_1}z_{j,a_2}z_{\kappa,a_3} \prod_{p}z_{a_p,a_{p+1}}}
    \times   \mathrm{PT}(i,j,\kappa^\prime) \frac{\mathrm{Pf} \mathbf{A}_{(j,n)\cup(n,i)}}{\prod_{q}z_{b_q, b_{q+1}}}
    ,
\end{aligned} \\
&
\label{eq2:YMScase3}
\begin{aligned}
    (3): \qquad\qquad &\frac{ \mathrm{Pf}^\prime \mathbf{A}_n}{z_{i,b_1} z_{j,b_2}z_{k,a_1}\prod_{p}z_{a_p,a_{p+1}} \prod_{q}z_{b_q, b_{q+1}}} \to\\
    &\Delta_{i,j,\kappa}  \frac{\mathrm{Pf} \mathbf{A}_{\{i\} \cup (i,j)}}{z_{i,j}z_{j,\kappa}z_{\kappa,a_1} \prod_{p}z_{a_p,a_{p+1}}}
    \times \Delta_{i,j,\kappa '}  \frac{\mathrm{Pf} \mathbf{A}_{(j,n)\cup(n,i)}}{z_{i,b_1}z_{j,b_2}z_{i,\kappa^\prime}z_{j,\kappa^\prime}\prod_{q}z_{b_q, b_{q+1}}}\\
    =&\Delta_{i,j,\kappa}  \frac{\mathrm{Pf} \mathbf{A}^{[j,\kappa]}_{\{i,j,\kappa\} \cup (i,j)}/z_{j,\kappa}}{z_{i,j}z_{\kappa,a_1} \prod_{p}z_{a_p,a_{p+1}}}
    \times \Delta_{i,j,\kappa '}   \frac{\mathrm{Pf} \mathbf{A}_{(j,n)\cup(n,i)}}{z_{i,b_1}z_{j,b_2}z_{i,\kappa^\prime}z_{j,\kappa^\prime}\prod_{q}z_{b_q, b_{q+1}}}
    ,
\end{aligned}
\end{align} 
\end{subequations}
}%
where we define $z_{i,a_1}\prod_{p}z_{a_p,a_{p+1}}:=z_{i,a_1}z_{a_2,a_3}\ldots z_{a_{|A|-1},a_{|A|}}$ and similar for those factors with a product over the index $q$. In the third line of these equations, we have rearranged them in a more suggesting form. For example, the third line of~\eqref{eq2:YMScase1} contains a factor that can be interpreted as a reduced Pfaffian (weighted by $\Delta_{i,j,\kappa}$),
\begin{equation}
   \Delta_{i,j,\kappa} \mathrm{Pf} \mathbf{A}^{[j,\kappa]}_{\{i,j,\kappa\} \cup (i,j)}/z_{j,\kappa}=  \mathrm{Pf}^\prime \mathbf{A}_{\{i,j,\kappa\} \cup (i,j)},
\end{equation}
and the corresponding denominator can be understood as a perfect matching of $\{i,j,\kappa\} \cup (i,j)$. In the split of the CHY integral, it will yield a pure current, and the remaining factor corresponds to a mixed current.

\subsection{CHY integrands and string correlators for gluons/gravitons}\label{sec_int of spin}
In this subsection, we discuss the splitting of CHY half-integrands and string correlators that are responsible for amplitudes with spinning particles, namely gluons and gravitons (as well as photons in Einstein-Maxwell or Born-Infeld theory), under the conditions in \eqref{eq_pol}. The basic ingredient is the reduced Pfaffian of $\mathbf{\Psi}_n$ matrix, an anti-symmetric $2n \times 2n$ matrix that gives gluon and graviton amplitudes via CHY formula~\cite{Cachazo:2013iea}, defined as
\begin{equation}
    \mathbf{\Psi}_n = 
    \left(\begin{array}{cc}
        \mathbf{A}_n & -\mathbf{C}_n^\mathrm{T}\\
        \mathbf{C}_n & \mathbf{B}_n
    \end{array}\right),
\end{equation}
where $\mathbf{B}_n, \mathbf{C}_n$ are $n \times n$ matrices involving the polarization vectors, whose components are given as
\begin{equation}
(\mathbf{B}_n)_{a b}=\begin{cases}
    \frac{2\epsilon_a \cdot \epsilon_b}{z_a-z_b} & a \neq b, \\
    0 & a=b,
\end{cases} \qquad
(\mathbf{C}_n)_{a b}=\begin{cases}
    \frac{2\epsilon_a \cdot p_b}{z_a-z_b} & a \neq b, \\
    \sum_{c\neq a}\frac{2\epsilon_a \cdot p_c}{z_a-z_c} & a=b.
\end{cases}
\end{equation}
The definition of the reduced Pfaffian of $\mathbf{\Psi}$ is the same as \eqref{eq:pfA},
\begin{equation}\label{eq_PfPsi}
    \operatorname{Pf}^{\prime} \mathbf{\Psi}_n :=\Delta_{i,j,k}  \frac{(-1)^{p+q}}{\left(z_p-z_q\right)} \operatorname{Pf}\left(\mathbf{\Psi}_n^{[p,q]}\right).
\end{equation}
Remarkably, we will see that not only does the $\mathbf{\Psi}_n$ matrix splits nicely (just as the $\mathbf{A}_n$ matrix above), but its splitting actually implies the splitting of superstring correlators~\cite{Mizera:2019gea}! This is possible due to a nice observation that the latter can be obtained by applying a certain differential operator to ${\rm Pf}' \mathbf\Psi_n$ (with respect to Mandelstam variables and Lorentz products of polarizations), thus the splitting of superstring correlator is a simple consequence of that of ${\rm Pf}' \mathbf\Psi_n$. Furthermore, we will also show that the bosonic string correlator splits nicely, under the same conditions \eqref{eq_pol}. 

\paragraph{${\rm Pf}' \mathbf{\Psi}_n$}
Since the reduced Paffian \eqref{eq_PfPsi} is independent of the choice of $p,q$, here we set $p=i, q=n$. To better illustrate the splitting, we exploit the permutation invariance of $\mathrm{Pf}' \mathbf{\Psi}^{[i,n]}_n$ (up to a sign) and arrange the components related to momenta and polarizations in set $(i,j)$ in the upper left block and those in set $(j,n)\cup (n,i)$ in the lower right block. Specifically, we reorganize the rows and columns of the matrix $\mathbf{\Psi}^{[i,n]}_n$ according to the ordering $\{ \alpha,\tilde \alpha,j,\beta,\tilde{\beta'}\}$, where $\alpha=(i,j)$, $\beta=(j,n)\cup(n,i)$, $\beta'=\{j\}\cup(j,i)\cup\{i\}$, $\tilde \alpha=(i+n,j+n)$ and $\tilde \beta'=\{j+n\}\cup(j+n,i+n)\cup\{i+n\}$ (note that we have used ordered subsets to denote the row and column indices of a matrix, {\it e.g.} ${\bf A}_{\alpha \times \alpha}$ means that the row and columns are both in the range $(i,j)$). Then the condition~\eqref{eq_pol} imposes
\begin{equation}\label{eq_split matrix Psi}
	\setlength{\arraycolsep}{0.1pt}
	\renewcommand{\arraystretch}{0.2}
	\mathbf{\Psi}^{[i,n]}_n\to
	\left(\begin{array}{cc|ccc}
			\begin{array}{c}\large{\mathbf{A}_{\alpha\times \alpha}}\end{array}&\begin{array}{c}\large{-\mathbf{C}^{\rm T}_{\alpha\times \tilde \alpha}}\end{array}&\begin{array}{c}\mathbf{A}_{i+1,j}\\ \vdots \\ \mathbf{A}_{j-1,j}\end{array}&\large{0}&\large{0}\\ 
			\begin{array}{c}\large{\mathbf{C}_{\tilde \alpha\times \alpha}}\end{array}&\begin{array}{c}\large{\mathbf{B}_{\tilde \alpha\times \tilde \alpha}}\end{array}&\begin{array}{c}\mathbf{C}_{i+1,j}\\ \vdots \\ \mathbf{C}_{j-1,j}\end{array}&\large{0}&\large{0}\\ \hline
			
			\begin{array}{c}\mathbf{A}_{j,i+1}\ldots \mathbf{A}_{j,j-1}\end{array}& \begin{array}{c}-\mathbf{C}^{\rm T}_{j,i+1} \ldots -\mathbf{C}^{\rm T}_{j,j-1}\end{array}&0&\begin{array}{c}\mathbf{A}_{j,j+1}\ldots \mathbf{A}_{j,i-1}\end{array}& \begin{array}{c}0-\mathbf{C}^{\rm T}_{j,j+1}\ldots-\mathbf{C}^{\rm T}_{j,i}\end{array}\\
			\large{0}&\large{0}&\begin{array}{c}\mathbf{A}_{i+1,j}\\ \vdots \\ \mathbf{A}_{j-1,j}\end{array}&\begin{array}{c}\large{\mathbf{A}_{\beta\times \beta}}\end{array}&\begin{array}{c}\large{-\mathbf{C}^{\rm T}}_{\beta\times \tilde \beta'}\end{array}\\
			\large{0}&\large{0}&\begin{array}{c}0\\ \mathbf{C}_{j+1,j}\\ \vdots \\ \mathbf{C}_{i,j}\end{array}&\begin{array}{c}\large{\mathbf{C}_{\tilde \beta'\times \beta}}\end{array}&\begin{array}{c}\large{\mathbf{B}_{\tilde \beta'\times \tilde \beta'}}\end{array}\\
	\end{array}\right).
\end{equation}
The condition $s_{a,b}=0$ implies that all components in the region $\alpha\times \beta$ are zero. Similarly, the condition~\eqref{eq_pol} implies some regions to become zero as shown in~\eqref{eq_split matrix Psi}.
And the condition $p_b\cdot\epsilon_{a}=0$ removes the components involving $(j,n)\cup(n,i)$ of the diagonal element $\mathbf{C}_{a,a}$ and the condition $p_a\cdot\epsilon_{b'}=0$ removes the components involving $(i,j)$ of the diagonal element $\mathbf{C}_{b', b'}$. Using the simple lemma in~\cite{Cachazo:2021wsz} again, we get the splitting of $\mathrm{Pf}'\mathbf{\Psi}_n$:
\begin{equation}\label{eq: splitting of PfPsi}
	\setlength{\arraycolsep}{1.5pt}
	\renewcommand{\arraystretch}{1.5}
	\begin{aligned}
		\Delta_{i,j,n}~\frac{1}{z_{i,n}}~\mathrm{Pf}\mathbf{\Psi}^{[i,n]}_n
		&\to
        \mathrm{Pf}\mathbf{\Psi}_{\{A\}}\times
		\Delta_{i,j,\kappa'}~\frac{1}{z_{i,\kappa'}}~\mathrm{Pf}\mathbf{\Psi}^{[i,\kappa']}_{\{j,B,i,\kappa'\}}\\[6pt]
        &=\mathrm{PT}(i,j,\kappa)\mathrm{Pf}\mathbf{\Psi}_{\{A\}}\times\mathrm{Pf}^\prime\mathbf{\Psi}_{\{j,B,i,\kappa'\}},
	\end{aligned}
\end{equation}
where $\mathbf{\Psi}_{\{i_1,i_2,\dots,i_m\}}$ denotes the submatrix with only columns and  rows in $\{i_1,i_2,\dots,i_m,i_1+n,i_2+n,\dots,i_m+n\}$ remaining. The right part obviously corresponds to the pure gluons/gravitons current (albeit with one off-shell leg $\kappa'$, which carries the polarization of leg $n$). As suggested in~\cite{Cheung:2017ems}, the left part also can be seen as the mixed one as follows:
\begin{equation}
    \mathrm{PT}(i,j,\kappa)\mathrm{Pf}\mathbf{\Psi}_{\{A\}}=\partial_{2\epsilon_i\cdot\epsilon_\kappa}\partial_{2\epsilon_j\cdot(p_i-p_\kappa)}\Bigg(\Delta_{i,j,\kappa}~\frac{1}{z_{i,\kappa}}~\mathrm{Pf}\mathbf{\Psi}^{[i,\kappa]}_{\{i,A,j,\kappa\}}\Bigg).
\end{equation}

\paragraph {Superstring correlator} Recall that the gauge-fixed suprstring correlator is defined  (see eq. (4.8) of~\cite{Mizera:2019gea}) as
\begin{equation}\label{eq_superstring correlator}
	\varphi_{n}^{\mathrm{type-I}}=\frac{\Delta_{i,j,n}}{z_{i_0, j_0}}\sum_{q=0}^{\lfloor n/2\rfloor -1}
	\sum_{\substack{\text{distinct}\\ \text{pairs}\\ \{i_l,j_l\}}}\prod_{l=1}^{q}\left(\frac{-2 \epsilon_{i_l}\cdot \epsilon_{j_l}}{\alpha'z_{i_l, j_l}^2}\right)\;\mathrm{Pf}{\mathbf\Psi}^{[i_0, j_0]}_{\{1,\dots,n\}\backslash\{i_1,j_1,\dots,i_q,j_q\}}.
\end{equation}
The second sum goes over all $q$ distinct unordered pairs $\{i_1,j_2\},\{i_2,j_2\},\dots,\{i_q,j_q\}$ of labels from the set $\{1,2,\dots,n\}\backslash\{i_0,j_0\}$, and $(i_0,j_0)$ is arbitrary. To demystify~\eqref{eq_superstring correlator} let us spell out a few leading terms
\begin{align*}
\varphi^{\mathrm{type-I}}_{n}&=\frac{\Delta_{i,j,n}}{z_{1,n}} \Bigg[\, \text{Pf}\ {\mathbf\Psi}_n^{[1,n]} -\sum_{2\leqslant i_1 < j_1 \leqslant n-1} \frac{2 \epsilon_{i_1} \cdot  \epsilon_{j_1}}{\alpha'z_{i_1,j_1}^2} \text{Pf}\ {\mathbf\Psi}^{[1,n]}_{\{1,\dots,n\}\backslash\{i_1,j_1\}} \\
&+\Bigg(\sum_{2 \leqslant i_1 < j_1 < i_2 < j_2 \leqslant n-1} \!{+}\! \sum_{2 \leqslant i_1 < i_2 < j_1 < j_2 \leqslant n-1}\Bigg) \frac{2 \epsilon_{i_1} \cdot  \epsilon_{j_1}}{\alpha'z_{i_1,j_1}^2} \frac{2 \epsilon_{i_2} \cdot  \epsilon_{j_2}}{\alpha' z_{i_2,j_2}^2} \text{Pf}\ {\mathbf\Psi}^{[1,n]}_{\{1,\dots,n\}\backslash\{i_1,j_1,i_2,j_2\}} + \ldots \Bigg],
\end{align*}
where we choose $(i_0,j_0)=(1,n)$. In order to simplify the formula of superstring correlators~\eqref{eq_superstring correlator}, we define an differential operator $\widehat{O}_{p,q}$:
\begin{equation}
    \widehat{O}_{p,q}:=-\frac{1}{\alpha^\prime}\epsilon_p\cdot\epsilon_q\partial_{\epsilon_p\cdot\epsilon_q}\partial_{s_{p,q}} \, .
\end{equation}
The differential operator $\widehat{O}_{p,q}$ acting on any $\mathrm{Pf} \mathbf{\Psi}$ removes the $\{p,q,p+n,q+n\}$ rows and columns and generates a overall factor:
\begin{equation}
    \widehat{O}_{p,q}\mathrm{Pf}\mathbf{\Psi}_{\{X,p,q\}}=-\frac{2\epsilon_p\cdot\epsilon_q}{\alpha^\prime z_{p,q}^2}~\mathrm{Pf}\mathbf{\Psi}_{\{X\}},
\end{equation}
where $X$ refers to other indices. Now we can represent each term of~\eqref{eq_superstring correlator} as certain differential operators acting on $\mathrm{Pf}^\prime\mathbf{\Psi}_n$:
\begin{equation}
    \varphi_{n}^{\mathrm{type-I}}=\frac{\Delta_{i,j,n}}{z_{i_0, j_0}}\sum_{q=0}^{\lfloor n/2\rfloor -1}
	\sum_{\substack{\text{distinct}\\ \text{pairs}\\ \{i_l,j_l\}}}\prod_{l=1}^{q}\widehat{O}_{i_l,j_l}\;\mathrm{Pf}{\mathbf\Psi}^{[i_0, j_0]}_{\{1,\dots,n\}}=\sum_{q=0}^{\lfloor n/2\rfloor -1}
	\sum_{\substack{\text{distinct}\\ \text{pairs}\\ \{i_l,j_l\}}}\prod_{l=1}^{q}\widehat{O}_{i_l,j_l}\;\mathrm{Pf}^\prime{\mathbf{\Psi}_n}.
\end{equation}
It's obvious that applying $\widehat{O}_{p,q}\widehat{O}_{q,r}$ must yield zero due to the multi-linearity in the polarization vectors, hence we can induce the following compact formula of superstring correlators
\begin{equation}
    \varphi^{\mathrm{type-I}}_{n}=\prod_{\{p,q\}\subset\{1,\dots,n\}\backslash\{i_0,j_0\}}(1+\widehat{O}_{p,q})~\mathrm{Pf}^\prime\mathbf{\Psi}_n.
\end{equation}
One can easily check that expanding the product of the operators generates all the terms that appear in the original formula~\eqref{eq_superstring correlator}. Based on this compact formula and the splitting of $\mathrm{Pf}'\mathbf{\Psi}_n$, the splitting of superstring correlators is manifest. First the condition $\epsilon_{a}\cdot\epsilon_{b'}=0$ induces $\widehat{O}_{a,b'}=0$, which split the operator products:
\begin{equation}
    \prod_{\{p,q\}\subset\{1,\dots,n\}\backslash\{i,n\}}(1+\widehat{O}_{p,q})\to
    \Bigg(\prod_{\{p,q\}\subset A}(1+\widehat{O}_{p,q})\Bigg) \Bigg(\prod_{\{p,q\}\subset\{j,B\}}(1+\widehat{O}_{p,q})\Bigg).
\end{equation}
where we choose $(i_0,j_0)=(i,n)$. Since all the $s_{p,q}$ surviving the differential operators are nonzero, we can straightforwardly split $\mathrm{Pf}^\prime\mathbf{\Psi}_n$ and get:
\begin{equation}
    \varphi^{\mathrm{type-I}}_{n}\to\Bigg(\prod_{\{p,q\}\subset A}(1+\widehat{O}_{p,q})~\mathrm{PT}(i,j,\kappa)\mathrm{Pf}\mathbf{\Psi}_{\{A\}}\Bigg) \Bigg(\prod_{\{p,q\}\subset\{j,B\}}(1+\widehat{O}_{p,q})~\mathrm{Pf}^\prime\mathbf{\Psi}_{\{j,B,i,\kappa'\}}\Bigg).
\end{equation}
The right part corresponds to a pure current while the left part to a mixed current, in the same way as we have demonstrated in the splitting of $\mathrm{Pf}^\prime\mathbf{\Psi}_n$. 
Finally, we can get the splitting of $\varphi^{\mathrm{type-I}}_{n}$:
\begin{equation}\label{eq: splitting of varphi}
	\varphi^{\mathrm{type-I}}_{n}\to\varphi^{\mathrm{type-I+color}}_{j-i+2}(i^{\phi},A,j^{\phi};\kappa^{\phi})\times \varphi^{\text{type-I}}_{n-j+i+1}(j,B,i;\kappa')\, ,
\end{equation}
where the off-shell leg $\kappa'$ also carries the polarization vector of leg $k$. For example, for $n=5$ and $i=1,j=3,k=5$, we have:
\begin{equation}
    \begin{aligned}
        \varphi^{\mathrm{type-I}}_{5}&=(1+\widehat{O}_{2,3}+\widehat{O}_{2,4}+\widehat{O}_{3,4})\frac{\Delta_{1,3,5}}{z_{1,5}}\mathrm{Pf}\mathbf{\Psi}^{[1,5]}_5\\[5pt]
        &=(1+\widehat{O}_{2,3})(1+\widehat{O}_{2,4})(1+\widehat{O}_{3,4})\mathrm{Pf}^\prime\mathbf{\Psi}_5\\[5pt]
        &\to \left(1\times(1+\widehat{O}_{3,4})\right)\left(\mathrm{PT}(1,3,\kappa)\mathrm{Pf}\mathbf{\Psi}_{\{2\}}\times \mathrm{Pf}^\prime\mathbf{\Psi}_{\{1,3,4,\kappa'\}}\right)\\[5pt]
        &=\mathrm{PT}(1,3,\kappa)\mathrm{Pf}\mathbf{\Psi}_{\{2\}}\times(1+\widehat{O}_{3,4})\mathrm{Pf}^\prime\mathbf{\Psi}_{\{1,3,4,\kappa'\}} \, .
    \end{aligned}
\end{equation}

\paragraph{Bosonic string correlator} The gauge-fixed bosonic string correlators for $n$-gluon scattering are given by:
\begin{equation}\label{eq_bosonic correlator}
	\mathcal{C}_n=\Delta_{i,j,k} \sum_{r=0}^{\lfloor n/2 \rfloor{+}1} \sum_{\{g,h\}, \{l\}} \prod_{s}^r W_{g_s, h_s} \prod_{t}^{n-2r} V_{l_t}, \quad V_{i}:= \sum_{j\neq i}^n \frac{\epsilon_i \cdot p_j}{z_{i,j}}, \quad W_{i,j}:=\frac{\epsilon_i \cdot \epsilon_j}{\alpha' z_{i,j}^2},
\end{equation}
where we have a summation over all partitions of $\{1,2,\ldots, n\}$ into $r$ pairs $\{g_s, h_s\}$ and $n-2r$ singlets $l_t$, each summand given by the product of $W$'s and $V$'s. For instance, the $4$-point correlator is
\begin{equation}
    \mathcal{C}_4= V_1 V_2 V_3 V_4+(V_1 V_2 W_{3,4}+\text{perm.})+(W_{1,2}W_{3,4}+\text{perm.}) \, .
\end{equation}
Now let us impose the splitting conditions~\eqref{eq_pol}, which enforce:
\begin{equation}
	W_{a,b'}=0,\quad  V_{a}= \sum_{c \neq a, c\notin B} \frac{\epsilon_{a} \cdot p_c}{z_{a,c}},\quad V_{b'}= \sum_{c \neq b', c\notin A} \frac{\epsilon_{b'} \cdot p_c}{z_{b',c}}.
\end{equation}
Therefore, the polarization of $A$ and $B'$ completely decouple, and the summations in $V_a, V_{b'}$ only involve $A\cup \{i,j,\kappa\}$ or $B \cup \{i,j,\kappa'\}$ (with $\kappa,\kappa'$ missing since we have fixed $z_{\kappa},z_{\kappa'}\to \infty$), respectively.
As a consequence, the bosonic string correlator behaves as:
\begin{equation}\label{eq: splitting of calC_n}
	\begin{aligned}
		\mathcal{C}_n (1,2,\dots,n)&\to \mathrm{PT}(i,j,\kappa)\left(\sum_{r=0}^{\lfloor |A|/2 \rfloor{+}1} \sum_{\{g,h\}, \{l\} \in A} \prod_{s}^r W_{g_s, h_s} \prod_{t}^{|A|-2r} V_{l_t}\right) \\
		& \times
		\Delta_{i,j,\kappa'}\left(\sum_{r=0}^{\lfloor |B'|/2 \rfloor{+}1} \sum_{\{g,h\}, \{l\} \in B'} \prod_{s}^r W_{g_s, h_s} \prod_{t}^{|B'|-2r} V_{l_t}\right)\\
		&=\frac{1}{\Delta_{i,j,\kappa}} \mathrm{PT}(i,j,\kappa)\mathcal{C}_{j-i-1}(A)\times \mathcal{C}_{n-j+i+1}(j,B,i;\kappa')\,,
	\end{aligned}
\end{equation}
where in the last line we obtain the product of two string correlators: one corresponds to mixed amplitude with $A$ to be gluons and $\{i,j, \kappa\}$ to be $\phi$'s~\cite{Schlotterer:2016cxa,He:2018pol,He:2019drm}, another corresponds to pure gluon amplitude with external legs $B \cup \{i,j,\kappa'\}$. Note we have an extra factor $1/\Delta_{i,j,\kappa}$ for the mixed current since we absorb a $\Delta_{i,j,\kappa}$ in the definitions of PT$(i,j,\kappa)$ and $\mathcal{C}_{j-i-1}(A)$. For example, for $n=4$ and $i=1,j=3,k=4$, we have 
\begin{equation}
    \begin{aligned}
        {\cal C}_{4}(1,2,3,4)&=\Delta_{1,3,4}[V_1 V_2 V_3 V_4+(V_1 V_2 W_{3,4}+\text{perm.})+(W_{1,2}W_{3,4}+\text{perm.})]\\[5pt]
        &=\Delta_{1,3,4}V_2[(V_1V_3V_4+(V_4W_{1,3}+\text{perm.})]+\Delta_{1,3,4}[W_{1,2}(\star)+W_{2,3}(\star\star)+W_{2,4}(\star\star\star)]\\[5pt]
        &\to \mathrm{PT}(1,3,\kappa)V_2 \times \Delta_{1,3,\kappa'}[V_1 V_3 V_{\kappa'}+(V_{\kappa'}W_{1,3}+\text{perm.})]\\[5pt]
        &= \frac{1}{\Delta_{1,3,\kappa}}\mathrm{PT}(1,3,\kappa)\mathcal{C}_1(2)\times \mathcal{C}_3(3,1;\kappa')\,,
    \end{aligned}
\end{equation}
where the terms denoted by ``$\star$'' is not important since $W_{1,2}=W_{2,3}=W_{2,4}=0$.

\section{Splittings of string amplitudes} \label{sec:stringsplit}
In this section, we apply the splitting of Koba-Nielsen factor, \eqref{eq: KN factor}, and that of various string correlators, to derive splitting of string amplitudes of stringy $\phi^3$ models and their deformations~\cite{Arkani-Hamed:2023lbd,Arkani-Hamed:2023mvg,Arkani-Hamed:2023swr,Arkani-Hamed:2023jry,Arkani-Hamed:2024nhp}, and those in superstring and bosonic string theories. 

\subsection{Splitting of stringy $\phi^3$ amplitudes}
We begin with the simplest string amplitudes with Parke-Talyor factors only. Let us first illustrate how the integration domain for open-string integrals split. The definition of stringy $\phi^3$ amplitudes (also known as $Z/J$ integrals in the literature~\cite{Mafra:2011nw,Mafra:2016mcc,Schlotterer:2018zce}) reads:
\begin{equation}
     Z_{\alpha|\beta} := \int_{D(\alpha)} \dd \mu_n^{\mathbb{R}}\, \mathrm{PT}(\beta)\,, \quad J_{\alpha|\beta} := \int_{\mathbb{C}^n} \dd \mu_n^{\mathbb{C}}\, \mathrm{PT}(\alpha)\mathrm{PT}(\beta)\,,
\end{equation}
where the open string integration domain is
\begin{equation}
    \mathcal{D}(\alpha):=\left\{\left(z_1, z_2, \ldots, z_n\right) \in \mathbb{R}^n \mid-\infty<z_{\alpha_1}<z_{\alpha_2}<\ldots<z_{\alpha_n}<\infty\right\}\,,
\end{equation}

For the $2$-split of (stringy) $\phi^3$ amplitudes, we need to specify the orderings. Without loss of generality, we choose one ordering to be the canonical ordering $\mathbb{I}$. Note we have also chosen $k=n$ and $i<j{-}1$; $A=(i,j):=\{i{+}1, \ldots, j{-}1\}, B=(j,n)\cup(n,i):=\{j{+}1, \ldots, n{-}1, 1, \ldots, i{-}1\}$.
For open-string integral $	 Z_{\mathbb{I}|\alpha}$, the integration domain $\mathcal{D}(\mathbb{I})$ also splits nicely under gauge fixing $z_{i}=0,z_{j}=1,z_{k}=z_{n}=\infty$.
\begin{equation}\label{eq: splitting of domain}
    \begin{aligned}
        \mathcal{D}(\mathbb{I})&=\left\{\left(z_1, z_2, \ldots, z_n\right)/(z_{i},z_{j},z_{n}) \in \mathbb{R}^{n-3} \mid-\infty<\ldots<z_{i-1}<0<\ldots<z_{j-1}<1<\ldots<z_{n-1}<\infty\right\}\\
        &=\left\{\left(z_{i},z_{i+1},\ldots,z_{j-1}, z_{j},z_{\kappa}\right)/(z_{i},z_{j},z_{\kappa}) \in \mathbb{R}^{n-3} \mid-\infty<0<z_{i+1}<\ldots<z_{j-1}<1<\infty\right\}\\
        &\times\left\{\left(z_1, \ldots,z_{i},z_{j},\ldots, z_{n-1},z_{\kappa'}\right)/(z_{i},z_{j},z_{\kappa'}) \in \mathbb{R}^{n-3} \mid-\infty<\ldots<z_{i-1}<0<1<\ldots<z_{n-1}<\infty\right\}\\
        &=\mathcal{D}_{(i,j)}\times \mathcal{D}_{(j,n)\cup(n,i)}\,,
    \end{aligned}
\end{equation}
where $\mathcal{D}_{(i,j)}$ and $\mathcal{D}_{(j,n)\cup(n,i)}$ are the domains with gauge fixing  $z_{i}=0,z_{j}=1,z_{\kappa}=z_{\kappa^{\prime}}=\infty$.

On the other hand, as we have discussed in section~\ref{sec:PT split}, the integration measure and the Parke-Talyor factors also splits correctly, therefore the result of open/closed string integrals under the splitting kinematics reads:
\begin{equation}
    \begin{aligned}
         Z_{\mathbb{I}|\alpha} &\to \int_{\mathcal{D}_{(i,j)}\times \mathcal{D}_{(j,n)\cup(n,i)}} d \mu^{\mathbb{R}}_{j-i+2} (i,A,j; \kappa) d\mu^{\mathbb{R}}_{n-j+i+1} (j,B,i; \kappa') \mathrm{PT}_{(i,j)} \times \mathrm{PT}_{(j,n)\cup(n,i)}\\
        &=\int_{\mathcal{D}_{(i,j)}} d \mu^{\mathbb{R}}_{j-i+2} (i,A,j; \kappa)  \mathrm{PT}_{(i,j)} \int_{ \mathcal{D}_{(j,n)\cup(n,i)}} d\mu^{\mathbb{R}}_{n-j+i+1} (j,B,i; \kappa')\mathrm{PT}_{(j,n)\cup(n,i)}\\
        &\equiv 	\mathcal{J}^{\phi^3,\mathbb{R}}(i,A,j,\kappa|i,\alpha(A),j,\kappa)\times 	\mathcal{J}^{\phi^3,\mathbb{R}}(j,B\cup \kappa',i|j,\alpha(B\cup \kappa'),i)\,,
    \end{aligned}
\end{equation}
where we define $B\cup \kappa'=\{j{+}1, \ldots, n{-}1, \kappa',1, \ldots, i{-}1\}$, and the $\alpha(A)$ denotes the permutation ordering $A$ according to $\alpha$. We also define the shorthand notation $ \mathrm{PT}_{(i,j)}\equiv\mathrm{PT}(i,(i,j),j,\kappa)$, and $ \mathrm{PT}_{(j,n)\cup(n,i)}\equiv \mathrm{PT}(j, (j,n), \kappa', (n,i), i)$. 
And the (stringy) current is given by
\begin{equation}
    \mathcal{J}^{\phi^3,\mathbb{R}}_{n}(\alpha|\beta)=\int_{\mathcal{D}(\alpha)} d \mu^{\mathbb{R}}_{n} (\beta; \kappa)  \mathrm{PT}(\beta)\,.
\end{equation}

Similarly the $2$-split of closed string integrals $J_{\alpha|\beta}$ reads
\begin{equation}\label{eq:J-int-2split}
    \begin{aligned}
       J_{\alpha|\beta} &\to \int_{\mathbb{C}^{j-i+2}\times \mathbb{C}^{n-j+i+1}} d \mu^{\mathbb{C}}_{j-i+2} (i,A,j; \kappa) d\mu^{\mathbb{C}}_{n-j+i+1} (j,B,i; \kappa') \mathrm{PT}^{\alpha}_{(i,j)}  \mathrm{PT}^{\alpha}_{(j,n)\cup(n,i)}\mathrm{PT}^{\beta}_{(i,j)}  \mathrm{PT}^{\beta}_{(j,n)\cup(n,i)}\\\
        &=\int_{\mathbb{C}^{j-i+2}} d \mu^{\mathbb{C}}_{j-i+2} (i,A,j; \kappa)  \mathrm{PT}^{\alpha}_{(i,j)} \mathrm{PT}^{\beta}_{(i,j)} \int_{ \mathbb{C}^{n-j+i+1}} d\mu^{\mathbb{C}}_{n-j+i+1} (j,B,i; \kappa')\mathrm{PT}^{\alpha}_{(j,n)\cup(n,i)}\mathrm{PT}^{\beta}_{(j,n)\cup(n,i)}\\
        &\equiv 	\mathcal{J}^{\phi^3,\mathbb{C}}(i,\alpha(A),j,\kappa|i,\beta(A),j,\kappa)\times 	\mathcal{J}^{\phi^3,\mathbb{C}}(j,\alpha(B\cup \kappa'),i|j,\beta(B\cup \kappa'),i)\,,
    \end{aligned}
\end{equation}
where $ \mathrm{PT}_{(i,j)}^{\alpha}\equiv\mathrm{PT}(i,\alpha(i,j),j,\kappa)$, and $ \mathrm{PT}_{(j,n)\cup(n,i)}^{\alpha}\equiv \mathrm{PT}(j, \alpha(j,n), \kappa', \alpha(n,i), i)$, and  the closed (stringy) currents are defined as
\begin{equation}
    \mathcal{J}^{\phi^3,\mathbb{C}}_{n}(\alpha|\beta)=\int_{\mathbb{C}^n} d \mu^{\mathbb{C}}_{n} (\alpha; \kappa) \mathrm{PT}(\alpha) \mathrm{PT}(\beta)\,.
\end{equation}

As it is shown in~\cite{Cao:2024gln}, the deformed Parke-Taylor factor introduced in~\cite{Arkani-Hamed:2023swr,Arkani-Hamed:2023jry,Arkani-Hamed:2024nhp} also splits nicely, which implies the splitting of NLSM and YMS amplitudes with certain flavor pairs under the low field theory limit. We do not repeat the discussions here, and we will rather derive the splitting of these amplitudes directly from their CHY formulas (including the splitting of the YMS amplitudes with more general flavor pairs) in section~\ref{sec_splitParticle}. 

\subsection{Splitting of superstring and bosonic string amplitudes}
Now we move on to the bosonic string and superstring amplitudes defined in sec~\ref{sec_int of spin} and derive the splittings of both open- and closed-string cases.

\paragraph{Open-string} For bosonic string, since the domain $\mathcal{D}(\mathbb{I})$ and the correlator $\mathcal{C}_n$ split as~\eqref{eq: splitting of domain} and~\eqref{eq: splitting of calC_n}, bosonic string amplitudes with the canonical ordering split as: 
\begin{equation}
	\begin{aligned}
		\int_{\mathcal{D}(\mathbb{I})} d\mu_n^{\mathbb{R}} \mathcal{C}_n\to
		&\int_{\mathcal{D}_{(i,j)}} d \mu^{\mathbb{R}}_{j-i+2} (i,A,j; \kappa)  
		\frac{1}{\Delta_{i,j,\kappa}} \mathrm{PT}(i,j,\kappa)\mathcal{C}_{j-i-1}(A)\times\\
		&\int_{ \mathcal{D}_{(j,n)\cup(n,i)}} d\mu^{\mathbb{R}}_{n-j+i+1} (j,B,i; \kappa')
		\mathcal{C}_{n-j+i+1}(j,B,i;\kappa').
	\end{aligned}
\end{equation}
Hence the general 2-split of bosonic string amplitudes reads:
\begin{equation}\label{eq: splitting of bosonic string}
	\begin{aligned}
		\mathcal{M}^{\text{bosonic}}_{n}(\alpha)\to
		\mathcal{J}^{\text{bosonic}+\text{color}}(i^\phi,\alpha(A),j^\phi;\kappa^\phi)\times
		\mathcal{J}^{\text{bosonic}}_{\mu}(j,\alpha(B\cup \kappa'),i)\epsilon_n^\mu,
	\end{aligned}
\end{equation}
where the superscript ``bosonic$+$color'' denotes the mixed current with 3 $\phi^3$ scalar.

For superstring amplitudes, as we have discussed in subsection~\ref{sec_int of spin}, the superstring correlators $\varphi^{\mathrm{type-I}}_{n}$ split as~\eqref{eq: splitting of varphi}, thus the splitting of the superstring amplitudes with the canonical ordering is given by:
\begin{equation}
	\begin{aligned}
		\int_{\mathcal{D}(\mathbb{I})} d\mu_n^{\mathbb{R}}~\varphi^{\text{type-I}}_{n}\to
		&\int_{\mathcal{D}_{(i,j)}} d \mu^{\mathbb{R}}_{j-i+2} (i,A,j; \kappa)  ~\varphi^{\text{type-I}+\text{color}}_{j-i+2}(i^{\phi},A,j^{\phi};\kappa^{\phi})\times\\
		&\int_{ \mathcal{D}_{(j,n)\cup(n,i)}} d\mu^{\mathbb{R}}_{n-j+i+1} (j,B,i; \kappa')
		~\varphi^{\text{type-I}}_{n-j+i+1}(j,B,i;\kappa').
	\end{aligned}
\end{equation}
For general ordering, say $\alpha$, the splitting can be easily generalized:
\begin{equation}\label{eq: splitting of superstring}
	\begin{aligned}
		\mathcal{M}^{\text{type-I}}_{n}(\alpha)\to
		\mathcal{J}^{\text{type-I}+\text{color}}(i^\phi,\alpha(A),j^\phi;\kappa^\phi)\times
		\mathcal{J}^{\text{type-I}}_{\mu}(j,\alpha(B\cup \kappa'),i)\epsilon_n^\mu.
	\end{aligned}
\end{equation}

 Let us present a 5-point example for the superstring amplitudes with $i=1,j=3,k=5$,
\begin{equation}\label{eq: example for superstring}
	\begin{aligned}
		\mathcal{M}^{\text{type-I}}_5(1,2,3,4,5)\to
		&\frac{8\Gamma \left(1+\alpha^\prime s_{1,2}\right) \Gamma \left(1+\alpha^\prime s_{2,3}\right)}{\alpha^\prime\Gamma \left(1+\alpha^\prime s_{1,2}+\alpha^\prime s_{2,3}\right)} \mathcal{A}^{\mathrm{YM}+\phi^3}(1^\phi,2,3^\phi;\kappa^\phi)\times\\
		&\frac{8\Gamma \left(1-\alpha^\prime  s_{1,4}-\alpha^\prime  s_{3,4}\right) \Gamma \left(1+\alpha^\prime  s_{3,4}\right)}{\alpha^\prime  \Gamma \left(1-\alpha^\prime  s_{1,4}\right)}\mathcal{A}^\mathrm{YM}(1,3,4;\kappa'),
	\end{aligned}
\end{equation}
which is exactly~\eqref{eq: splitting of superstring}.

\paragraph{Closed-string} 
The splitting of open-string amplitudes immediately indicates that closed-string theory splits in exactly the same way since they share the same ingredients. There are two ways of imposing the condition~\eqref{eq_pol} to $\epsilon,\tilde{\epsilon}$, which correspond to two different splitting behaviours:
\begin{equation}
\label{eq_split1 of gr in string}
    \mathcal{M}^{\mathrm{closed}}_n\xrightarrow[\tilde{\epsilon}_a \cdot \tilde{\epsilon}_{b'},p_a \cdot \tilde{\epsilon}_{b'},\tilde{\epsilon}_a \cdot p_b=0]{p_a\cdot p_b,\epsilon_a \cdot \epsilon_{b'},p_a \cdot \epsilon_{b'},\epsilon_a \cdot p_b=0}  \mathcal{J}^{\mathrm{mixed}}(i^\phi, A, j^\phi; \kappa^\phi)\times\mathcal{J}_{\mu\nu}(i,j,B;\kappa')\epsilon_n^\mu\tilde\epsilon_n^\nu\,,
\end{equation}
\begin{equation}\label{eq_split2 of gr in string}
    \mathcal{M}^{\mathrm{closed}}_n\xrightarrow[\tilde{\epsilon}_{a'} \cdot \tilde{\epsilon}_{b},p_a \cdot \tilde{\epsilon}_{b},\tilde{\epsilon}_{a'} \cdot p_b=0]{p_a\cdot p_b,\epsilon_a \cdot \epsilon_{b'},p_a \cdot \epsilon_{b'},\epsilon_a \cdot p_b=0}
    \mathcal{J}^{\mathrm{EYM}}_{\nu}(i^g, A, j^g; \kappa^g)\tilde\epsilon_n^\nu\times\mathcal{J}_{\mu}^{\mathrm{EYM}}(i^g,j^g,B;\kappa^{g,\prime})\epsilon_n^\mu\,.
\end{equation}
Let us illustrate~\eqref{eq_split1 of gr in string} via an 5-point example with $i=1,j=3,k=5$, $A=\{2\}$ and $B=\{4\}$. We use the KLT relation~\cite{Kawai:1985xq} to compute the result instead of directly perform the modulus squared integrals:
\begin{equation}
	\mathcal{M}^{\mathrm{type-II}}_5=\sum_{\alpha\in X,\beta\in Y}\mathcal{M}^{\mathrm{type-I}}_5(\alpha)m_{\alpha'}^{-1}(\alpha|\beta)\mathcal{\tilde M}^{\mathrm{type-I}}_5(\beta).
\end{equation}
In a choice of orderings $X=\{(1,2,3,4,5),(1,2,4,3,5)\}$ and $Y=\{(1,3,2,5,4), (1,4,2,5,3)\}$ we have:
\begin{equation}
	\mathcal{M}^{\mathrm{type-II}}_5=\mathcal{M}^{\mathrm{type-I}}_5(1,2,3,4,5)\sin(\pi\alpha' s_{2,3})\sin(\pi\alpha' s_{4,5})\mathcal{\tilde M}^{\mathrm{type-I}}_5(1,3,2,5,4)+(3\leftrightarrow4).
\end{equation}
where the term $(3\leftrightarrow4)$ would vanish after splitting due to $s_{2,4}=0$. Since $\mathcal{M}^{\mathrm{type-I}}_5(1,3,2,5,4)$ is not a standard ordering in our choice of $A$ and $B$, we expand it to the follow BCJ basis:
\begin{equation}
	\mathcal{M}^{\mathrm{type-I}}_5(1,3,2,5,4)=\frac{\sin(\pi\alpha's_{1,2})\mathcal{M}^{\mathrm{type-I}}_5(1,2,3,5,4)+\sin(\pi\alpha's_{2,4})\mathcal{M}^{\mathrm{type-I}}_5(1,4,2,5,3)}{\sin(-\pi\alpha'(s_{1,2}+s_{2,3}))}.
\end{equation}
Then we have:
\begin{equation}
	\mathcal{M}^{\mathrm{type-II}}_5=\frac{\sin(\pi\alpha' s_{2,3})\sin(\pi\alpha' s_{4,5})\sin(\pi\alpha's_{1,2})}{\sin(-\pi\alpha'(s_{1,2}+s_{2,3}))}\mathcal{M}^{\mathrm{type-I}}_5(1,2,3,4,5)\mathcal{\tilde M}^{\mathrm{type-I}}_5(1,2,3,5,4)+\sin(\pi\alpha's_{2,4})(\star),
\end{equation}
where terms denoted by $\sin(\pi\alpha's_{2,4})(\star)$ vanishes since $s_{2,4}=0$. Using the results of the splitting of type-I string amplitudes, we get:
\begin{equation}
	\begin{aligned}
		\mathcal{M}^{\mathrm{type-II}}_5\to&\frac{\sin(\pi\alpha' s_{2,3})\sin(\pi\alpha's_{1,2})}{\sin(-\pi\alpha'(s_{1,2}+s_{2,3}))}\mathcal{J}^{\mathrm{type-I+color}}(1^\phi,2,3^\phi,\kappa^\phi)\mathcal{\tilde J}^{\mathrm{type-I+color}}(1^\phi,2,3^\phi,\kappa^\phi)\\
		&\times\sin(-\pi\alpha' (s_{1,4}+s_{3,4}))\mathcal{J}^{\mathrm{type-I}}_{\mu}(1,3,4,\kappa')\epsilon_n^\mu\mathcal{\tilde J}^{\mathrm{type-I}}_{\nu}(1,3,\kappa',4)\tilde\epsilon_n^\nu\\[5pt]
		=&\sin(-\pi\alpha's_{1,2})\mathcal{J}^{\mathrm{type-I+color}}(1^\phi,2,3^\phi,\kappa^\phi)\mathcal{\tilde J}^{\mathrm{type-I+color}}(1^\phi,2,\kappa^\phi,3^\phi)\\
		&\times\sin(\pi\alpha' (s_{1,4}+s_{3,4}))\mathcal{J}^{\mathrm{type-I}}_{\mu}(1,3,4,\kappa')\epsilon_n^\mu\mathcal{\tilde J}^{\mathrm{type-I}}_{\nu}(1,3,\kappa',4)\tilde\epsilon_n^\nu\\[5pt]
		=&\mathcal{J}^{\mathrm{type-II+color}}(1^\phi,2,3^\phi;\kappa^\phi)\times\mathcal{J}^{\mathrm{type-II}}_{\mu\nu}(1,3,4;\kappa')\epsilon_n^\mu\tilde\epsilon_n^\nu,
	\end{aligned}
\end{equation}
where we have used BCJ relation $\sin(\pi\alpha' s_{2,3})\mathcal{M}^{\mathrm{type-I}}(1,2,3,4)+\sin(\pi\alpha' (s_{1,2}+s_{2,3}))\mathcal{M}^{\mathrm{type-I}}(1,2,4,3)=0$ in the first equality.

Let us end this section with some comments on the NLSM. As suggested in~\cite{Dong:2024klq}, general dimensional reductions of bosonic and superstring theories give stringy completion of NLSM. We find such stringy NLSM model also splits under the kinematic locus~\eqref{eq_splitKin}, which can be made obvious under specific choice of dimensional reductions at the integrand level. Similar argument also holds for the stringy model of sGal. For these theories, we do not present the details of their splitting behavior here but will be focus on their field theory limit using the CHY formula in section~\ref{sec_splitParticle} .

\section{Splittings of particle amplitudes}
\label{sec_splitParticle}
In this section we consider the splittings of scalar and gluon/graviton 
amplitudes in the field-theory limit via their CHY formulae. In particular, in the some kinematical loci, the YMS/EMS/DBI amplitudes split into lower point currents times an object defined by its CHY formula. We also show that the naive splittings of scaffolded YMS/EMS amplitudes and the YM/GR ones are related by gauge transformations.

\subsection{Splitting of scalar amplitudes}
For scalar amplitudes, the CHY half integrands we need to consider here are the PT$(\alpha)$, det$'\mathbf{A}_n$ and  Pf$'\mathbf{A}_n/(z_{t_1 ,t_2}\ldots z_{t_{n-1},t_{n}})$. These building blocks can be combined into CHY integrands of amplitudes with color ordering, flavor pairs which require the split kinematics to be compatible with, or totally permutation invariant one~\cite{Cachazo:2014xea}. In this subsection, we will illustrate all these types of amplitudes via the Bi-adjoint $\phi^3$, NLSM, YMS and sGal, respectively. 

\subsubsection{Bi-adjoint $\phi^3$} \label{sec: splitting of Bi-adjoint phi3}
The half-integrands of the bi-adjoint $\phi^3$ amplitudes $\mathcal{A}^{\phi^3}(\alpha|\beta)$ are PT$(\alpha)$ and PT$(\beta)$, which are proved to 2-split in~\eqref{eq: splitting of PT}. In fact, the amplitudes splits the same as the J-integral~\eqref{eq:J-int-2split} since it is the low energy limit of the latter:
\begin{equation} \label{eq: phi3 split}
    \mathcal{A}^{\phi^3}(\alpha_1,\ldots,\alpha_n|\beta_1,\ldots,\beta_n) \to {\cal J}^{\phi^3}(i,\alpha(A),j,\kappa|i,\beta(A),j,\kappa)\;  {\cal J}^{\phi^3}(i,\alpha(B),j; \kappa^{\prime}|i,\beta(B),j; \kappa^{\prime})\,,
\end{equation}
where the $\alpha$ and $\beta$ are permutations of the set $\{1,\ldots,n\}$. As we mentioned before, $\alpha$ and $\beta$ must be compatible with~\eqref{eq_splitKin}, {\it i.e.} the elements in $A$ and $B$ must be separated by $i$ and $j$. On the RHS, each of the two resulting currents contains one off-shell leg, and has the rigorous definition in the CHY formalism~\cite{Naculich:2015zha}. As a result of the off-shellness, a physical pole of the current could be either massive or massless, depending on the factorization at this pole. When one of the two factors consists of only massless particles, the pole is massless, otherwise it is not.

Let us consider a $6$-point bi-adjoint $\phi^3$ amplitude ${\cal A}^{\phi^3}(\alpha|\beta)$, and choose $i=1,j=4,k=6$, $A=\{2,3\}$, and $B=\{5\}$ to construct the 2-split kinematics. 
In other words, we set $s_{2,5}=s_{3,5}=0$.
If both orderings $\alpha$ and $\beta$ are canonical, we observe the expected splitting,
\begin{align}
\label{eq_6ptphi3}
    {\cal A}^{\phi^3}(1,\ldots,6|1,\ldots,6)
    \xrightarrow[]{s_{a,b}=0}& 
    \left(\frac{1}{s_{1,2} s_{1,2,3}}+\frac{1}{s_{2,3} s_{1,2,3}}+\frac{1}{s_{2,3} s_{2,3,4}}+\frac{1}{s_{3,4} s_{2,3,4}}+\frac{1}{s_{1,2} s_{3,4}} \right) \times 
    \left(\frac{1}{s_{4,5}}+\frac{1}{s_{5,6}} \right)
    \nonumber \\ 
    =&\ {\cal J}^{\phi^3}(1,2,3,4; \kappa | 1,2,3,4; \kappa) \times {\cal J}^{\phi^3}(1,4,5; \kappa^\prime | 1,4,5; \kappa^\prime),
\end{align}
where $p_\kappa = -\sum_{\alpha=1}^{4} p_\alpha$, and $p_{\kappa^\prime} = -p_1 -p_4-p_5$, which restore momentum conservation.
The latter four-point current with an off-shell leg reads
\begin{equation}
    {\cal J}^{\phi^3} (1,4,5;\kappa^\prime) 
    = \frac{1}{s_{4,5}} + \frac{1}{s_{5,\kappa'}- p_{\kappa^\prime}^2}
    = \frac{1}{s_{4,5}} + \frac{1}{s_{5,6}},
\end{equation}
where the pole of $s_{4,5}$ is massless, because the particles $4,5$ are massless. But the pole associated with $s_{5,\kappa^\prime}$ is massive, under the 2-split kinematics $s_{2,5}=s_{3,5}=0$, it simplifies to $s_{5,6}$~\cite{Naculich:2015zha}.
The same simplification occurs for the five-point current, leaving no massive pole in $\ {\cal J}^{\phi^3}(1,2,3,4; \kappa | 1,2,3,4; \kappa) $ \eqref{eq_6ptphi3}.

\subsubsection{NLSM} \label{sec: splitting of NLSM}
Let us begin with the NLSM amplitudes, unlike the stringy model, the half integrand det$'A_n$ is the reduced determinant of a perfect anti-symmetric matrix whose splitting can be easily derived as we have seen in Section~\ref{sec: splitting of matrix A}. Combining with the result of the PT factor, we see that the splitting is given by:
\begin{equation}
\begin{aligned}
 \mathrm{PT}(1,2,\ldots,n)\; \mathrm{det}^\prime \mathbf{A}_n & \to  \mathrm{PT}(i,\ldots,j,\kappa)\; \mathrm{PT}(j,\kappa) \mathrm{det} \mathbf{A}_{\{i\} \cup (i,j)}\\
&  \mathrm{PT}(j,\ldots,n-1,\kappa^\prime,1,\ldots,i)\; \mathrm{PT}(i,j,\kappa^\prime) \mathrm{det} \mathbf{A}_{(j,n)\cup(n,i)}.
\end{aligned}
\end{equation}
where we have fixed the punctures $z_i\to 0, z_j\to 1, z_n,z_\kappa,z_{\kappa^\prime}\to \infty$ and assumed $|(i,j)|$ to be odd. Here on the RHS, after performing the integration we obtain the product of a $(|(i,j)|+3)$-point pure pion current (or equivalently, a current with $|(i,j)|+1$ pions with $2$ bi-adjoint scalar $j,\kappa$) and a $(|(i,j)|+3)$-point mixed current~\cite{Cachazo:2016njl} with pions $(j,n)\cup(n,i)$ and bi-adjoint scalar $i,j,\kappa^\prime$, {\it i.e.},
\begin{equation} \label{eq: NLSM split}
    \mathcal{A}^\mathrm{NLSM}(1,2,\ldots,n) \to {\cal J}^\mathrm{NLSM}(i,A,j;\kappa)\;  {\cal J}^{\mathrm{NLSM}+\phi^3}(i^\phi,B,j^\phi; \kappa^{\prime\, \phi}).
\end{equation}
One can easily derive the splittings for even $|(i,j)|$ in a similar way.

Let us present a 6-point example with $i=3,j=5,k=6$, therefore we have $A=\{4\}$, $B=\{1,2\}$ and the splitting reads
\begin{equation}
\begin{aligned}
\mathcal{A}^\mathrm{NLSM}(1,2,\ldots,6)=&s_{1,2}+s_{1,6}+s_{2,3}+s_{3,4}+s_{4,5}+s_{5,6}-\frac{\left(s_{1,2}+s_{2,3}\right) \left(s_{4,5}+s_{5,6}\right)}{s_{1,2,3}}\\
&-\frac{\left(s_{2,3}+s_{3,4}\right) \left(s_{1,6}+s_{5,6}\right)}{s_{2,3,4}}-\frac{\left(s_{1,2}+s_{1,6}\right) \left(s_{3,4}+s_{4,5}\right)}
{s_{3,4,5}}\\
\xrightarrow[]{s_{1,4}=s_{2,4}=0}& (s_{3,4}+s_{4,5}) \times (1-\frac{s_{1,2}}{s_{1,2,3}}-\frac{s_{1,2}}{s_{3,4,5}}-\frac{s_{2,3}}{s_{1,2,3}}-\frac{s_{2,3,4,5}}{s_{3,4,5}})\\
=& {\cal J}^\mathrm{NLSM}(3,4,5;\kappa)\;  {\cal J}^{\mathrm{NLSM}+\phi^3}(1,2,3^\phi,5^\phi;\kappa^{\prime\, \phi}).
\end{aligned}
\end{equation}
For another explicit examples with $i=1,j=4,k=6$, we have
\begin{equation} \label{eq: NLSM example 6pt}
\begin{aligned}
\mathcal{A}^\mathrm{NLSM}(1,2,\ldots,6)
\xrightarrow[]{s_{2,5}=s_{3,5}=0}& (1-\frac{s_{1,2}}{s_{1,2,3}}-\frac{s_{2,3}}{s_{1,2,3}}-\frac{s_{2,3}}{s_{2,3,4}}-\frac{s_{3,4}}{s_{2,3,4}}) \times s_{1,5}\\
=&   {\cal J}^{\mathrm{NLSM}+\phi^3}(1^\phi,2,3,4^\phi;\kappa^{\phi})\; {\cal J}^\mathrm{NLSM}(1,4,5;\kappa^\prime),
\end{aligned}
\end{equation}
Let us spell out two more examples at 10-point: 
\begin{equation} \label{eq: NLSM example 10pt1}
\begin{aligned}
    \mathcal{A}^\mathrm{NLSM}(1,2,\ldots,10)& \xrightarrow[]{i=2,j=5,k=10} {\cal J}^{\mathrm{NLSM}+\phi^3}(2^\phi,3,4,5^\phi; \kappa^\phi)\; {\cal J}^{\mathrm{NLSM}}(1,2,5,6,7,8,9; \kappa^{\prime}),
\end{aligned}
\end{equation}
\begin{equation} \label{eq: NLSM example 10pt2}
\begin{aligned}
    \mathcal{A}^\mathrm{NLSM}(1,2,\ldots,10)& \xrightarrow[]{i=2,j=6,k=10} {\cal J}^\mathrm{NLSM}(2,3,4,5,6; \kappa)\; {\cal J}^{\mathrm{NLSM}+\phi^3}(1,2^\phi,6^\phi,7,8,9; \kappa^{\prime\, \phi}).
\end{aligned}
\end{equation}

\subsubsection{Special Galileon} 
\label{subsubsec:sGal}
For the sGal, we now have both the left and right CHY half-integrands as $\det' \mathbf{A}_n$. In analogy with the NLSM cases, for odd $|(i,j)|$ the splitting reads:
\begin{equation}
\begin{aligned}
  (\mathrm{det}^\prime \mathbf{A}_n)^2 \to&   \mathrm{PT}(j,\kappa)  (\mathrm{det}' \mathbf{A}_{(i,j)})^2 \times  (\mathrm{PT}(i,j,\kappa^\prime) \mathrm{det} \mathbf{A}_{(j,n)\cup(n,i)})^2,
\end{aligned}
\end{equation}
which is integrated to be:
\begin{equation}
    \mathcal{A}^\mathrm{sGal}_n \to {\cal J}^\mathrm{sGal}(i,A,j; \kappa)\;  {\cal J}^ {\mathrm{sGal}+\phi^3}(i^\phi,B,j^\phi; \kappa^{\prime\, \phi} ).
\end{equation}
Now crucially, the amplitudes of the sGal are permutation invariant, so any permutation of the kinematic conditions would induce corresponding splittings.
Let us give an explicit examples at $n=6$ with $i=3,j=5,k=6$:
\begin{equation}
\begin{aligned}
\mathcal{A}^\mathrm{sGal}_6
\xrightarrow[]{s_{1,4}=s_{2,4}=0}& s_{3,4} s_{4,5} (s_{3,4}+s_{4,5}) \times \frac{-s_{1,2}}{s_{1,2,3} s_{1,2,5} s_{3,4,5}}  \Big((s_{2,3} s_{3,4,5}+s_{1,2,3} (s_{2,3,4,5}-s_{3,4,5})){}^2\\
&-(s_{2,3} s_{3,4,5}+s_{1,2,3} (s_{2,3,4,5}-s_{3,4,5})) s_{1,2}^2)-s_{1,2} (s_{3,4,5} s_{2,3}^2-s_{3,4,5} (2 s_{1,2,3}+s_{3,4,5}) s_{2,3}\\
&+s_{1,2,3} (s_{3,4,5}-s_{2,3,4,5}) (s_{1,2,3}+s_{3,4,5}-s_{2,3,4,5})\Big)\\
=& {\cal J}^\mathrm{sGal}(3,4,5;\kappa)\;  {\cal J}^{\mathrm{sGal}+\phi^3}(1,2,3^\phi,5^\phi;\kappa^{\prime\, \phi}).
\end{aligned}
\end{equation}
Similarly, for $i=1,j=4,k=6$ we have
\begin{equation} \label{eq: sGal example 6pt}
\begin{aligned}
\mathcal{A}^\mathrm{sGal}_6
\xrightarrow[]{s_{2,5}=s_{3,5}=0}&  \frac{s_{2,3}}{s_{1,2,3} s_{1,4,5} s_{2,3,4}}\Big(\left(s_{3,4} s_{1,2,3}+\left(s_{1,2}-s_{1,2,3}\right) s_{2,3,4}\right) s_{2,3}^2\\
&+\left(s_{1,2,3} s_{3,4}^2-s_{1,2,3} \left(s_{1,2,3}+2 s_{2,3,4}\right) s_{3,4}+\left(s_{1,2}-s_{1,2,3}\right) \left(s_{1,2}-s_{1,2,3}-s_{2,3,4}\right) s_{2,3,4}\right) s_{2,3}\\
&-\left(s_{3,4} s_{1,2,3}+\left(s_{1,2}-s_{1,2,3}\right) s_{2,3,4}\right){}^2 \Big) \times s_{1,5} s_{4,5} (s_{1,5}+s_{4,5})\\
=&   {\cal J}^{\mathrm{sGal}+\phi^3}(1^\phi,2,3,4^\phi;\kappa^{\phi})\; {\cal J}^\mathrm{sGal}(1,4,5;\kappa^\prime). 
\end{aligned}
\end{equation}
Of course, one can easily go beyond 6-point and consider $n=10$:
\begin{equation} \label{eq: sGal example 10pt1}
\begin{aligned}
    \mathcal{A}^\mathrm{sGal}_{10}& \xrightarrow[]{i=3,j=6,k=10} {\cal J}^{\mathrm{sGal}+\phi^3}(3,4,5,6; \kappa^\phi)\; {\cal J}^{\mathrm{sGal}}(1,2,3,6,7,8,9; \kappa^{\prime}),
\end{aligned}
\end{equation}
\begin{equation} \label{eq: sGal example 10pt2}
\begin{aligned}
    \mathcal{A}^\mathrm{sGal}_{10}& \xrightarrow[]{i=3,j=7,k=10} {\cal J}^\mathrm{sGal}(3,4,5,6,7; \kappa)\; {\cal J}^{\mathrm{sGal}+\phi^3}(1,2,3^\phi,7^\phi,8,9; \kappa^{\prime\, \phi}).
\end{aligned}
\end{equation}

\subsubsection{YMS, EMS and DBI} \label{sec:YMSfac}
The key ingredient needed for YMS amplitudes is the half integrand Pf$^\prime \mathbf{A}_n/\left(z_{t_1,t_2}\ldots z_{t_{n-1},t_n}\right)$ whose behavior under the splitting kinematics has been discussed in Section~\ref{sec: splitting of matrix A}. Note that the conclusion we derive here for YMS amplitudes also applies to Einstein-Maxwell-scalar and Dirac-Born-Infeld in a similar way.  Let us take another half integrand to be $\mathrm{PT}(1,2,\ldots,n)$ and assume the split kinematics are also compatible with the ordering, using the results in~\eqref{eq:YMSallcases}, it is direct to show that the amplitudes behave as:
\begin{equation} \label{eq:YMSAmpcase1}
\begin{aligned}
    &\mathcal{A}^{\rm YMS}(\{i,a_1\},\{a_2,a_3\},\ldots,\{b_1,b_2\},\ldots,\{j,k\}) \to\\
    &{\cal J}^{\rm YMS}(\{i,a_1\},\{a_2,a_3\},\ldots,\{j,\kappa\}) \times   {\cal J}^{\mathrm{YMS}+\phi^3}(\{b_1,b_2\},\ldots,\{i,j,\kappa^\prime \}),
\end{aligned}
\end{equation}
\begin{equation} \label{eq:YMSAmpcase2}
\begin{aligned}
    &\mathcal{A}^{\rm YMS}(\{i,a_1\},\{j,a_2\},\{k,a_3\},\{a_4,a_5\},\ldots,\{b_1,b_2\},\ldots,\{b_{|B|-1},b_{|B|}\}) \to\\
    &{\cal J}^{\rm YMS}(\{i,a_1\},\{j,a_2\},\{\kappa,a_3\},\ldots,\{a_{|A|-1},a_{|A|}\})  \times   {\cal J}^{\mathrm{YMS}+\phi^3}(\{b_1,b_2\},\ldots,\{i,j,\kappa^\prime \}),
\end{aligned}
\end{equation}
\begin{equation} \label{eq:YMSAmpcase3}
\begin{aligned}
    &\mathcal{A}^{\rm YMS}(\{k,a_1\},\{a_2,a_3\},\ldots,\{i,b_1\},\{j,b_2\},\ldots,\{b_{|B|-1},b_{|B|}\}) \to\\
    &{\cal J}^{\rm YMS}(\{\kappa,a_1\},\{a_2,a_3\},\ldots,\{i,j\})  
     \times    \int d \mu^\mathrm{CHY}_{|B|+3} \Delta_{i,j,\kappa'} \frac{\mathrm{Pf} \mathbf{A}_{(j,n)\cup(n,i)}\mathrm{PT}(j,\ldots,n-1,\kappa',1,\ldots,i)}{z_{i,b_1}z_{j,b_2}z_{i,\kappa^\prime}z_{j,\kappa^\prime}\prod_{p}z_{b_p,b_{p+1}}} ,
\end{aligned}
\end{equation}
where we have 3 cases correspond to three types of perfect matching in~\eqref{eq:YMScase1}$\sim$\eqref{eq:YMScase3}, respectivley. For the first and second cases, the results are the product of an even point YMS current and an odd point mixed current with $3$ bi-adjoint scalars; for the third case, we obtain a YMS current and an object defined by this CHY formula.

Let us illustrate this with some explicit examples at 6 points. For the case~\eqref{eq:YMSAmpcase1}, we choose the perfect matching $\{1,2\},\{3,4\},\{5,6\}$ with the split kinematics given by $i=3,j=5,k=6$, then we have:
\begin{equation}
\begin{aligned}
\mathcal{A}^\mathrm{YMS}(\{1,2\},\{3,4\},\{5,6\})
\xrightarrow[]{s_{1,4}=s_{2,4}=0}& (1+\frac{s_{4,5}}{s_{3,4}}) \times(\frac{s_{2,3}}{s_{1,2} s_{1,2,3}}+\frac{s_{2,3,4,5}}{s_{1,2} s_{3,4,5}}-\frac{1}{s_{1,2}}+\frac{1}{s_{1,2,3}}+\frac{1}{s_{3,4,5}})\\
=& {\cal J}^\mathrm{YMS}(\{3,4\},\{5,\kappa\})\;  {\cal J}^{\mathrm{YMS}}(\{1,2\},\{3,5,\kappa^\prime\}).
\end{aligned}
\end{equation}
Note that there is no non-trivial example at 6-point corresponds to~\eqref{eq:YMSAmpcase2} since it requires $A$ or $B$ to contain at least 3 elements. Now let us focus on the case~\eqref{eq:YMSAmpcase3}, if we take $i=1,j=3,k=6$ for $n=6$, there are 3 permutation-inequivalent perfect matchings: $\{\{1,2\},\{3,4\},\{5,6\}\}$, $\{\{1,2\},\{3,5\},\{4,6\}\}$, $\{\{2,6\},\{1,5\},\{3,4\}\}$. For the first two cases, the splittings read:
\begin{equation} 
\begin{aligned}
    &\mathcal{A}^\mathrm{YMS}(\{1,2\},\{3,4\},\{5,6\}) \xrightarrow[]{s_{2,4}=s_{2,5}=0}
    (1+\frac{s_{2,3}}{s_{1,2}}) \times (\frac{1}{s_{4,5,6}}+\frac{1}{s_{3,4,5}}+\frac{s_{4,5}}{s_{5,6}s_{4,5,6}}++\frac{s_{4,5}}{s_{3,4}s_{5,6}}+\frac{s_{4,5}}{s_{3,4}s_{3,4,5}}) \\
    &={\cal J}^{\mathrm{YMS}}(\{1,2\},\{3,\kappa\})    \times    \int d \mu^\mathrm{CHY}_{5} \Delta_{i,j,\kappa'}  \mathrm{PT}(3,4,5,6,\kappa^\prime) \frac{\mathrm{Pf} \mathbf{A}_{\{4,5\}}}{z_{3,4}z_{5,6}z_{4,\kappa^\prime}z_{5,\kappa^\prime}},
\end{aligned}
\end{equation}
\begin{equation} 
\begin{aligned}
    &\mathcal{A}^\mathrm{YMS}(\{1,2\},\{3,5\},\{4,6\}) \xrightarrow[]{s_{2,4}=s_{2,5}=0}
    (1+\frac{s_{2,3}}{s_{1,2}}) \times -(\frac{1}{s_{4,5,6}}+\frac{1}{s_{3,4,5}}) \\
    &={\cal J}^{\mathrm{YMS}}(\{1,2\},\{3,\kappa\})    \times    \int d \mu^\mathrm{CHY}_{5} \Delta_{i,j,\kappa'}  \mathrm{PT}(3,4,5,6,\kappa^\prime) \frac{\mathrm{Pf} \mathbf{A}_{\{4,5\}}}{z_{3,5}z_{4,6}z_{4,\kappa^\prime}z_{5,\kappa^\prime}},
\end{aligned}
\end{equation}
where in the above examples we have treated $i=1$ as the off-shell leg $\kappa'$ in the explicit CHY formula on the RHS. For the last perfect matching that leads to an exceptional object, it is natural to keep $k=6$ as the off-shell leg:
\begin{equation} 
\begin{aligned}
    &\mathcal{A}^\mathrm{YMS}(\{2,6\},\{1,5\},\{3,4\}) \xrightarrow[]{s_{2,4}=s_{2,5}=0}
    -1 \times (\frac{1}{s_{3,4,5}}+\frac{s_{4,5}}{s_{3,4} s_{3,4,5}}) \\
    &={\cal J}^{\mathrm{YMS}}(\{\kappa,2\},\{1,3\})    \times   \underbrace{\int d \mu^\mathrm{CHY}_{5} \Delta_{i,j,\kappa'}  \mathrm{PT}(1,3,4,5,\kappa^\prime) \frac{\mathrm{Pf} \mathbf{A}_{\{4,5\}}}{z_{1,5}z_{3,4}z_{1,\kappa^\prime}z_{3,\kappa^\prime}}}_{(*)} .
\end{aligned}
\end{equation}
It is noteworthy that the exceptional object can be extracted from a pure YM amplitude via differential operators, {\it e.g.}
\begin{equation}
    (*) = \partial_{\epsilon_1 \cdot \epsilon_5} \partial_{\epsilon_3 \cdot \epsilon_4} (\partial_{\epsilon_{\kappa'} \cdot p_{1} } -\partial_{\epsilon_{\kappa'} \cdot p_3}) \mathcal{A}^\mathrm{YM}(1,3,4,5,\kappa').
\end{equation}
Importantly, the combination of the differential operators we have used here are {\it not} those\footnote{The differential operator that reduces the spin of the particles in $\alpha$ takes the general form: $\hat{T}(\alpha):=\partial_{\epsilon_{\alpha_1} \cdot \epsilon_{\alpha_r}}  (\partial_{\epsilon_{\alpha_2} \cdot p_{\alpha_1} } -\partial_{\epsilon_{\alpha_2} \cdot p_{\alpha_r}}) (\partial_{\epsilon_{\alpha_3} \cdot p_{\alpha_2} } -\partial_{\epsilon_{\alpha_3} \cdot p_{\alpha_r}})\ldots  (\partial_{\epsilon_{\alpha_{r-1}} \cdot p_{\alpha_{r-2}} } -\partial_{\epsilon_{\alpha_{r-1}} \cdot p_{\alpha_r}})$ for an ordered set $\alpha$ with length $r$.} in~\cite{Cheung:2017ems}, which confirms $(*)$ is not the amplitude considered in~\cite{Cachazo:2014xea,Cheung:2017ems}. It is argued in~\cite{Cheung:2017ems} that the operators we used here can preserve the momentum conservation but break the gauge invariance. However, for YMS amplitudes we consider here, one finally obtains a scalar amplitude via the operators, therefore the violation of the gauge invariance is not problematic. Let us end this subsection with examples for $n=10$:

\begin{equation} \label{eq: YMS example 10pt1}
\begin{aligned}
    &\mathcal{A}^{\rm YMS}(\{1,2\},\{3,4\},\ldots,\{9,10\}) \xrightarrow[]{i=3,j=7,k=10}\\
    & \mathcal{J}^{\rm YMS}(\{3,4\},\{5,6\},\{7,\kappa\}) \mathcal{J}^{ \mathrm{YMS}+\phi^3}(\{7,8\},\{1,2\},\{3,9,\kappa'\}) \, ,
\end{aligned}
\end{equation}
\begin{equation}  \label{eq: YMS example 10pt2}
\begin{aligned}
    &\mathcal{A}^{\rm YMS}(\{1,2\},\{3,4\},\ldots,\{9,10\}) \xrightarrow[]{i=4,j=7,k=10}\\
    &  \mathcal{J}^{ \mathrm{YMS}+\phi^3}(\{5,6\},\{4,7,\kappa\}) \mathcal{J}^{\rm YMS}(\{1,2\},\{3,4\},\{7,8\},\{9,\kappa'\}) \, .
\end{aligned}
\end{equation}

\subsection{Splittings of gluon/graviton amplitudes} \label{sec_split of spin particles}
Now let us consider the splitting of gluon and graviton amplitudes. Crucially, in addition to the splitting kinematic~\eqref{eq_splitKin} for Mandelstam variables that leads to the splitting of the measure, we need to further impose~\eqref{eq_pol} that involves the polarization vectors to ensure the splitting of integrands. Combining~\eqref{eq: splitting of PT} and~\eqref{eq: splitting of PfPsi}, it is straightforward to derive
\begin{equation}\label{eq:splitting of YM}
\mathcal{A}^{\rm YM}(1,2,\ldots,n)  \xrightarrow[]{p_a\cdot p_b,\epsilon_a \cdot \epsilon_{b'},p_a \cdot \epsilon_{b'},\epsilon_a \cdot p_b=0} {\cal J}^{\mathrm{YM}+\phi^3} (i^\phi, A, j^\phi; \kappa^\phi) \times {\cal J}^\mathrm{YM}_\mu (j, B, i; \kappa') \epsilon_n^\mu\,,  
\end{equation}
for $a\in A, b \in B$ and $b' \in B \cup \{i,j,n\}$.

For instance, at $7$ points, we pick $i=1, j=4, k=7$, and set
\begin{equation}
    \label{eq_7ptYMkin}
    s_{a,b}
    = \epsilon_{a} \cdot \epsilon_{b^\prime} 
    = p_{a} \cdot \epsilon_{b^\prime} 
    = \epsilon_{a} \cdot p_{b} 
    = 0, \quad
    \text{ for } \quad
    a \in \{2,3\},\  
    b \in \{5,6\},\
    b^\prime \in \{5,6\}\cup \{1,4,7\},
\end{equation}
such that the YM amplitude becomes
\begin{equation}
\label{eq_7ptYM2split}
\begin{aligned}
    {\cal A}^\mathrm{YM}(1,2,3,4,5,6,7)& \xrightarrow[]{i=1,j=4,k=7} 
    {\cal J}^{\mathrm{YM}+\phi^3} (1^\phi,2,3,4^\phi; \kappa^\phi)\; 
    {\cal J}^{\mathrm{YM}}_{\mu} (1,4,5,6; \kappa^{\prime}) \epsilon_{7}^{\mu} ,
\end{aligned}
\end{equation}
where $\epsilon_7^\mu$ should be reinterpreted as associated with the off-shell momentum $\kappa^\prime$. 

Surprisingly, for YM, an even simpler split kinematic with set $B$ being empty is possible.
This is non-trivial since one still needs to decouple the polarizations of $i,j,k$ from the particles in set $A$ to observe the 2-split behavior.
It becomes evident even at $4$ points, where, with $\epsilon_2 \cdot \epsilon_{b^\prime}=p_2 \cdot \epsilon_{b^\prime}=0$ for $b^\prime \in \{1,3,4\}$ the YM amplitude splits as
\begin{equation}\label{eq:ym4pt2split}
\begin{aligned}
    {\cal A}^\mathrm{YM}(1,2,3,4) \xrightarrow[]{i=1,j=3,k=4}
    &\left(-\frac{p_3\cdot \epsilon _2}{s_{2,3}}
    +\frac{p_1\cdot \epsilon _2}{s_{1,2}}\right)
    \left(
      \epsilon_1 \!\cdot\! \epsilon _3\, p_{3}^{\mu}
      + \epsilon_3 \!\cdot\! p_1 \, \epsilon_{1}^{\mu}
      + \epsilon_1 \!\cdot\! p_{\kappa^{\prime}} \, \epsilon_{3}^{\mu} \right)
    \epsilon_{4,\mu} \\
    &=
    {\cal J}^{\mathrm{YM}+\phi^3}(1^\phi,2,3^\phi; \kappa^\phi)\; 
    {\cal J}^{\mathrm{YM}}_{\mu}(1,3; \kappa^{\prime}) \epsilon_{4}^{\mu}.
\end{aligned}
\end{equation}

Interestingly, for graviton amplitudes where we have two copies of the polarization vetors, namely $\epsilon, \tilde{\epsilon}$, we can impose conditions \eqref{eq_pol} separately on $\epsilon$ and $\tilde{\epsilon}$ which lead to different splittings. In one choice, $\epsilon, \tilde{\epsilon}$ for particles $i,j,k$ live in the same current and we obtain the product of a mixed current of graviton and $\phi^3$ and a pure graviton one:
\begin{equation}\label{eq:splitting of GR 1}
\mathcal{A}^{\rm GR}(1,2,\ldots,n)  \xrightarrow[\tilde{\epsilon}_a \cdot \tilde{\epsilon}_{b'},p_a \cdot \tilde{\epsilon}_{b'},\tilde{\epsilon}_a \cdot p_b=0]{p_a\cdot p_b,\epsilon_a \cdot \epsilon_{b'},p_a \cdot \epsilon_{b'},\epsilon_a \cdot p_b=0} {\cal J}^{\mathrm{GR}+\phi^3} (i^\phi, A, j^\phi; \kappa^\phi) \times {\cal J}^\mathrm{GR}_{\mu\nu} (j, B, i; \kappa') \epsilon_n^\mu \tilde{\epsilon}_n^\nu\,. 
\end{equation}
In another choice, $\epsilon, \tilde{\epsilon}$ for particles $i,j,k$ are distributed into different currents and we obtain two EYM currents with $i,j,k$ to be gluons in both currents, (the gluon is denoted as $i^{g},j^{g},\kappa^{g}$):
\begin{equation}\label{eq:splitting of GR 2}
\mathcal{A}^{\rm GR}(1,2,\ldots,n)  \xrightarrow[\tilde{\epsilon}_{a'} \cdot \tilde{\epsilon}_{b},p_a \cdot \tilde{\epsilon}_{b},\tilde{\epsilon}_{a'} \cdot p_b=0]{p_a\cdot p_b,\epsilon_a \cdot \epsilon_{b'},p_a \cdot \epsilon_{b'},\epsilon_a \cdot p_b=0} {\cal J}^{\rm EYM}_{\nu} (i^g, A, j^g; \kappa^g) \tilde{\epsilon}_n^\nu \times {\cal J}^\mathrm{EYM}_{\mu} (j^g, B, i^g; \kappa^{g,\prime}) \epsilon_n^\mu \, ,
\end{equation}
where we have defined $a' \in A \cup \{i,j,k\}$.

Let us take a $7$-point amplitude as an explicit example.
If we assign both polarizations to the same side, \textit{i.e.}, \eqref{eq_7ptYMkin} applies identically to $\epsilon_\mu, \tilde{\epsilon}_\mu$, the amplitude splits in a similar way as the YM one~\eqref{eq_7ptYM2split},
\begin{equation}
\label{eq_6ptGR2split}
\begin{aligned}
    {\cal A}^\mathrm{GR}(1,2,3,4,5,6,7)& \xrightarrow[]{i=1,j=4,k=7} 
    {\cal J}^{\mathrm{GR}+\phi^3} (1^\phi,2,3,4^\phi; \kappa^\phi)\; 
    {\cal J}^{\mathrm{GR}}_{\mu\nu} (1,4,5,6; \kappa^{\prime}) \epsilon_{7}^{\mu} \tilde{\epsilon}_{7}^{\nu} ,
\end{aligned}
\end{equation}
where we note that the second term is pure GR.
Alternatively, if we adopt \eqref{eq_7ptYMkin} only for $\epsilon_\mu$, and enforce the following conditions for $\tilde{\epsilon}_\mu$,
\begin{equation}
    \label{eq_6ptGRkin}
     \tilde{\epsilon}_{b} \cdot \tilde{\epsilon}_{a^\prime}
    = p_{a} \cdot \tilde{\epsilon}_{b} 
    =  p_{b} \cdot \tilde{\epsilon}_{a^\prime}
    = 0, \quad
    \text{ for } \quad
    a \in \{2,3\},\  
    b \in \{5,6\},\
    a^\prime \in \{2,3\}\cup \{1,4,7\},
\end{equation}
then we obtain two mixed currents, each with three gluons and the remaining particles being gravitons,
\begin{equation}
\label{eq_6ptGR2split2}
\begin{aligned}
    {\cal A}^\mathrm{GR}(1,2,3,4,5,6,7)& \xrightarrow[]{i=1,j=4,k=7} 
    {\cal J}^{\mathrm{EYM}}_{\nu} (1^g,2,3,4^g; \kappa^g)\tilde\epsilon_{7}^{\nu} \; 
    {\cal J}^{\mathrm{EYM}}_{\mu} (1^g,4^g,5,6; \kappa^{g,\prime}) {\epsilon}_{7}^{\mu}.
\end{aligned}
\end{equation}

\subsection{Comments on the splittings of gluon amplitudes}
Let us briefly comment on two different ways of getting the splitting of the YM amplitudes: (1) the direct splitting of the $n$-point YM amplitudes under~\eqref{eq_splitKin} and~\eqref{eq_pol}; (2) the splitting of gluon amplitudes induced from scaffolded~\cite{Arkani-Hamed:2023mvg,Arkani-Hamed:2023swr} $2n$-point YMS amplitudes with flavor pairs $\{1,2\},\{3,4\},\ldots,\{2n-1,2n\}$ in~\eqref{eq:YMSAmpcase1} and \eqref{eq:YMSAmpcase2}. Note they are not generally the same, however, we will demonstrate their connections via an example.

We begin by analyzing the 2-split of the 10-point YMS amplitudes, the special kinematics with $i=3$, $j=10$, and $k=5$ reads:
\begin{equation}\label{eq:2splits10pt}
s_{1,4}=s_{2,4}=s_{1,6}=s_{2,6}=s_{1,7}=s_{2,7}=s_{1,8}=s_{2,8}=s_{1,9}=s_{2,9}=0 \, .
\end{equation}
Under these conditions, the 10-point amplitude splits into a 5-point current multiplied by an 8-point one. Now we take the scaffolding residues, {\it i.e.} the residues on $s_{1,2}=s_{3,4}=s_{5,6}=s_{7,8}=s_{9,10}=0$. Then we identify the polarization $\epsilon_i$ and momentum $k_i$ of the 5-point gluon amplitude to be $\epsilon_{i}=p_{2i-1}$, $k_{i}=p_{2i-1}+p_{2i}$ for $i = (1,2,3,4,5)$ (with $p_i$ representing momentum in the original 10-point amplitude). Hence conditions \eqref{eq:2splits10pt} transform into:
\begin{gather}
     \epsilon_{4}\cdot k_{1}=\epsilon_{5}\cdot k_{1}=\epsilon_{4}\cdot \epsilon_{1}=\epsilon_{5}\cdot \epsilon_{1}=k_{4}\cdot k_{1}=k_{4}\cdot \epsilon_{1}=0,\\
     \epsilon_{2}\cdot k_{1}=k_{2}\cdot k_{1},
     \quad \epsilon_{3}\cdot k_{1}=k_{3}\cdot k_{1},
     \quad \epsilon_{2}\cdot \epsilon_{1}=\epsilon_{2}\cdot k_{1},
     \quad \epsilon_{3}\cdot \epsilon_{1}=\epsilon_{3}\cdot k_{1}\,.
\end{gather}

It is worth noting that the first line of conditions is very similar to the 2-split kinematics for the 5-point gluon amplitude with $i=2$, $j=5$, and $k=3$. However, the second line differs. Nevertheless, we can perform a gauge transformation on $\epsilon_{2}$ and $\epsilon_{3}$, namely $\epsilon_{2}\to \epsilon_{2}-k_{2}$ and $\epsilon_{3}\to \epsilon_{3}-k_{3}$. With this gauge transformation, the second line of conditions becomes $\epsilon_{2}\cdot k_{1}=\epsilon_{3}\cdot k_{1}=\epsilon_{2}\cdot \epsilon_{1}=\epsilon_{3}\cdot \epsilon_{1}=0$. Now, these conditions exactly match our 2-split condition when $i=2$, $j=5$, and $k=3$. 

We conjecture this can be generalized to higher points: the splittings of gluon amplitudes in (1) and (2) are related by gauge transformations. We expect it also holds for the direct splitting of graviton amplitudes and the one obtained from the splitting of scaffolded EMS amplitudes.

\section{Implications of the splittings}\label{sec:implications}
In this section, we present several implications of our $2$-split behavior. Perhaps the most important one is the factorizations near zeros~\cite{Arkani-Hamed:2023swr} for which we will carefully present new results and examples for all theories we considered. Relatedly, we will consider a special ``skinny" case of the splitting and derive universal soft behavior from it, for gluons/gravitons and for Goldstone particles, respectively. Finally, we will comment on multi-splitting behavior which can be understood as further splitting our $2$-split results: in addition to the $3$-splits considered in~\cite{Cachazo:2021wsz}, we will also go to the extreme and consider the ``maximal splitting" where the amplitude becomes the product of four-point currents only.

\subsection{Factoriaztion near zeros}
\begin{figure}[tb]
    \centering
    \includegraphics[scale=0.45]{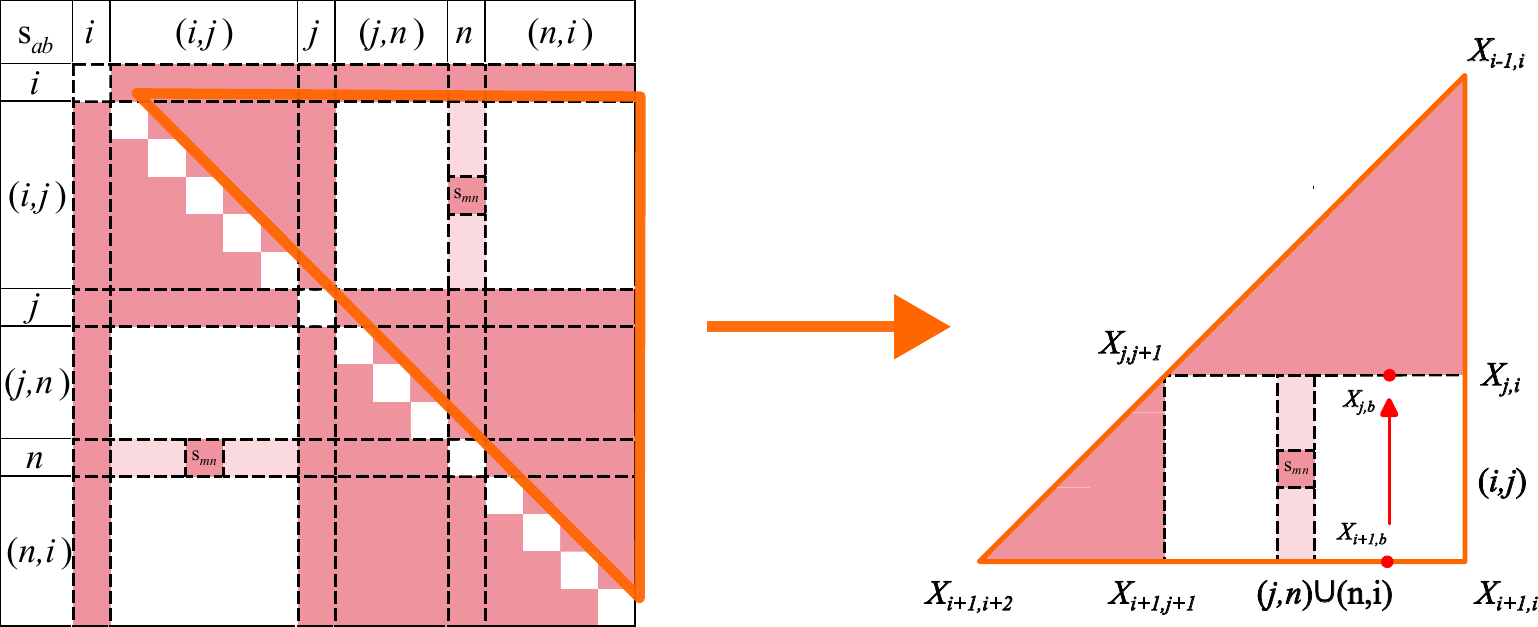}
    \caption{The Mandelstam matrix (left) and the kinematic mesh (right) for $2$-split kinematics: $s_{a,b}=0$ are denoted by blank entries, and the additional $s_{a,n}=0$ (except for $a=m$) are denoted by light pink entries, with $a \in A=(i,j):=\{i{+}1, \ldots, j{-}1\}, b\in B = (j,n)\bigcup(n,i) = \{j{+}1, \ldots, n{-}1, 1, \ldots, i{-}1\} $. Under the splitting, $X_{i,j}$ variables will be shifted as indicated by the arrow in the mesh (right), \textit{c.f.} \cite{Arkani-Hamed:2023swr}. }
    \label{fig_ManMatrix2mesh}
\end{figure}

For the relation of our 2-split and the so-called ``factorization near zeros"~\cite{Arkani-Hamed:2023swr}, note that in the special case when $A$ has only one particle, denoted as $m$, \eqref{eq_splitKin} corresponds to the factorization near ``skinny'' zero of~\cite{Arkani-Hamed:2023swr}: the amplitude factorizes into an $(n{-}1)$-point current (with on-shell legs $\{1,2,\ldots,n\}\backslash{\{k,m\}}$ and the off-shell leg $\kappa$), times a $4$-point function (with on-shell legs $i,j$ and two more off-shell legs), as well as a trivial $3$-point current. More generally, if we further set $s_{a,k}=0$ for all $a\in A$ {\it except for $a=m$}, then we expect a further split:
\begin{equation}
    {\cal S}_L(i, A, j; \kappa)\to {\cal S}_L(i, A\backslash\{m\}, j; \rho) + {\cal S}(i, \rho', j, \kappa),
\end{equation}
where we have used that $s_{a,\kappa}=0$ for $a \in A$ and $a\neq m$. Very nicely, this gives the familiar factorizations near zeros (since the four-point current always vanishes when we finally set $s_{m,k}=0$). Without loss of generality, we will choose $k=n$ and $i<j{-}1$, $A=(i,j):=\{i{+}1, \ldots, j{-}1\}, B=(j,n)\cup(n,i):=\{j{+}1, \ldots, n{-}1, 1, \ldots, i{-}1\}$; any 2-split kinematics can be obtained by relabelling. Note that those zeros of color-ordered amplitudes of (stringy) Tr~$\phi^3$, NLSM or YMS theory, correspond to $s_{a,b}=0$ for $i<a<j$ and $j<b\leq n$ (including $b=n$) and $1\leq b<i$, which is precisely given by a ``rectangle" in the mesh picture of associahedron (there are $n(n{-}3)/2$ such zeros in total); by excluding $s_{m, n}=0$ for some $i<m<j$ we recover the factorization near zero in the mesh.
The relationship between the Mandelstam matrix and the kinematic mesh is illustrated in Figure \ref{fig_ManMatrix2mesh}.
However, we will see that our result implies generalizations of such factorizations near zeros to amplitudes without color-ordering, such as closed-string amplitudes and those in special Galileon, Dirac-Born-Infeld and Einstein-Maxwell-scalar theories. 

Furthermore, we find that the shifts of the kinematic variables $X_{a,b}$ of the lower-point amplitudes is simply a consequence of the presence of massive external legs in the split kinematics. 
The key is to note that the $X_{a,b}$ variables should be reinterpreted with respect to the split lower-point amplitudes. 
For example, the amplitude corresponding to ${\cal S}_R(i,j,B; \kappa')$ depends on $X_{a,b}$'s in the upper triangle of the kinematics mesh.
Considering $X_{j,b}$ in the bottom boundary, if $1 \leq b < i$, we have
\begin{align}
    X_{j,b} \quad \to \quad& (p_j + \dots + p_{n-1} + p_{\kappa'} + p_1 + \dots + p_b)^2 \nonumber \\
    =& (p_j + \dots + p_{n-1} + (p_n + p_{i+1} + \dots + p_{j-1}) + p_1 + \dots + p_b)^2 = X_{i+1,b}.
\end{align}
And for $j\leq b <n$, there is no need to shift $X_{j,b}$ as it does not involve the massive leg $\kappa'$.
Analogously, the variables on the right boundary of the lower triangle $X_{a,j+1}$ is shifted to $X_{a,i}$ for $i<a<m$.
This is exactly the kinematic shift presented in \cite{Arkani-Hamed:2023swr}.

In the following, we will show factorization near zeros for all these string and particle amplitudes, which directly follows from our $2$-split. Of course the zeros of these amplitude then follow from the four-point function: the full amplitude vanishes if we further set $s_{m,n}=0$ for particle scattering and $s_{m,n}$ being any non-positive integer for string scattering. 

\subsubsection{Factoriaztion near zeros of scalar amplitudes}
Let us first consider scalar amplitudes in bi-adjoint $\phi^3$, NLSM/YMS, DBI/EMS and sGal. For the first three cases, the factorizations near zeros have already been studied in~\cite{Arkani-Hamed:2023swr} from their unified stringy integrals which was the motivation for all our studies. As it has been mentioned in~\cite{Cao:2024gln} (see also~\cite{Bartsch:2024ofb}),  exactly the same factorizations near zeros also apply to amplitudes without color ordering, and the general claim is that in any of these scalar theories:
\begin{itemize}
    \item The amplitude vanishes for $s_{a,b'}=0$ with $a\in A$ and $b'\in B':=B \cup \{k\}$. 
    \item The amplitude factorizes when we turn on $s_{k,m}\neq 0$, for any $m \in A$:
\begin{equation}
{\cal A}_n \to {\cal A}_4 \times {\cal J}(i, A\backslash\{m\}, j; \rho) \times {\cal J}(j, B, i; \rho')\,.
\end{equation}
\end{itemize}
Here let us consider sGal to be concrete: if both $|A|, |B|$ are even, these are sGal currents times ${\cal A}_4^{\phi^3}$; if both $|A|, |B|$ are odd, these are mixed currents with $3$ $\phi$'s, times ${\cal A}^{\rm sGal}_4=-s t (s+t)$ ($s:=s_{i,\kappa}$, $t:=s_{j,\kappa}$). In the following, we will investigate the factorizations near zeros for all these theories.

\paragraph{Bi-adjoint $\phi^3$}
For bi-adjoint $\phi^3$, the factorization near zeros can be achieved by the further conditions $s_{a,n}=0,a\in A\backslash\{m\}$ after the 2-split~\eqref{eq: phi3 split}. From the similar splitting of the half-integrand~\eqref{eq: splitting of PT} and the integration domain~\eqref{eq: splitting of domain}, the current $\mathcal{J}^{\phi^3}$ in~\eqref{eq: phi3 split} splits as 
\begin{equation}
   {\cal J}^{\phi^3}(i,\alpha(A),j,\kappa|i,\beta(A),j,\kappa)\xrightarrow[]{s_{a,n}=0, a\in A\backslash\{m\}} {\cal J}^{\phi^3}(\alpha_{m+1},\ldots,j,i,\ldots,\alpha_{m-1};\rho) {\cal J}^{\phi^3}(i,\rho^\prime,j,\kappa)\,,
\end{equation}
where we have omitted the second ordering in the two currents, and the factorization near zeros also demands the condition $m=\alpha_m=\beta_m$. The four-point current is 
\begin{equation} \label{eq: 4pt phi3 current}
    {\cal J}^{\phi^3}(i,\rho^\prime,j,\kappa)=\frac{1}{s_{i,\kappa}}+\frac{1}{s_{j,\kappa}}=\frac{-s_{m,n}}{s_{i,\kappa} s_{j,\kappa}}\,,
\end{equation}
where we should remark that this four-point current is generally ill-defined in the CHY formalism, because the CHY can only handle at most 3 off-shell legs~\cite{Naculich:2015zha}, whereas in the four-point kinematics, the amplitude/string integral can be retained by the only one-dimensional CHY/string integration. Therefore, the four-point current can always be defined by the one dimension integral from CHY/string, see also~\cite{Arkani-Hamed:2023swr}.

Let us continue the 6-point example in subsubsection~\ref{sec: splitting of Bi-adjoint phi3}.
For factorization near zeros, we further choose $m=2$, imposing the condition $s_{3,6} = 0$ (recall that $i=1,j=4,k=6$).
The amplitude factorizes as
\begin{equation}
    \begin{aligned}
        \eqref{eq_6ptphi3}
        \xrightarrow[]{s_{3,6}=0}& 
        \left(\frac{1}{s_{3,4}}+\frac{1}{s_{2,3}} \right) \times \left(\frac{1}{s_{1,2,3}}+\frac{1}{s_{2,3,4}} \right) \times 
        \left(\frac{1}{s_{4,5}}+\frac{1}{s_{5,6}} \right) \\
        =&\ {\cal J}^{\phi^3} (3,4,1; \rho | 3,4,1;\rho) 
        \times {\cal J}^{\phi^3} (1,\rho^\prime,4,\kappa | 1,\rho^\prime,4,\kappa) 
        \times {\cal J}^{\phi^3} (1,4,5;\kappa^\prime | 1,4,5;\kappa^\prime),
    \end{aligned}
\end{equation}
where $p_\rho = -p_1-p_3-p_4$, $p_{\rho^\prime} = -p_1-p_4-p_{\kappa}$, and ${\cal J}^{\phi^3} (1,\rho^\prime,4,\kappa | 1,\rho^\prime,4,\kappa)$ is the universal $4$-point object.

\paragraph{NLSM}
Now let us illustrate how the splitting is related to the factorizations near zeros for NLSM~\cite{Arkani-Hamed:2023swr}. In addition to the splitting kinematics which leads the 2-split in~\eqref{eq: NLSM split}, we further impose $s_{a,n}=s_{a,\kappa}=0, a\in A\backslash\{m\}$. It is important to note this kinematic locus for the set $\{i,A,j,\kappa\}$ is nothing but the condition~\eqref{eq_splitKin} with three special legs $i'=j,j'=i,k'=m$. Moreover, we are free to change the gauge choice of the punctures and the deleted columns and rows in the CHY integrand without changing the result of CHY integral that gives the current. Therefore, given~\eqref{eq: splitting of PT} and~\eqref{eq:A_split_odd}, it is easy to show
\begin{equation}
    {\cal J}^\mathrm{NLSM}(i,\ldots,j,\kappa) \xrightarrow[]{s_{a,n}=0,a\in A\backslash\{m\}} {\cal J}^{\mathrm{NLSM}+\phi^3}(m+1,\ldots,j^\phi,i^\phi,\ldots,m-1; \rho^\phi) {\cal J}^\mathrm{NLSM}(i,\rho^\prime,j,\kappa),
\end{equation}
for odd $|(i,j)|$, where we have the current of pure pions further splits into a mixed one and a 4-pion current with two off-shell legs. For even $|(i,j)|$,  the current of pure pions remains untouched, and using the results in section~\ref{sec: splitting of matrix A}, it can be shown that the mixed current behaves as:
\begin{equation}
    {\cal J}^{\mathrm{NLSM}+\phi^3}(i^\phi,\ldots,j^\phi,\kappa^\phi) \xrightarrow[]{s_{a,n}=0,a\in A\backslash\{m\}} {\cal J}^{\mathrm{NLSM}}(m+1,\ldots,j,i,\ldots,m-1; \rho) {\cal J}^{\phi^3}(i,\rho^\prime,j,\kappa),
\end{equation}
Crucially we now have a current of pure pions and a 4-point Tr~$\phi^3$ one with two off-shell legs, which is consistent with counting of mass dimension. Note for both cases, the 4-point current is proportional to $s_{m,n}$ on the support of the special kinematics~\eqref{eq_splitKin}:
\begin{equation}
    {\cal J}^\mathrm{NLSM}(i,\rho^\prime,j,\kappa)=s_{i,\kappa}+s_{j,\kappa}=-s_{m,n},
\end{equation}
similar to ${\cal J}^{\phi^3}(i,\rho^\prime,j,\kappa)$ as given in~\eqref{eq: 4pt phi3 current}. Therefore, if we further set $s_{m,n}=0$, the amplitudes vanish for both cases, this is exactly the zeros studied in~\cite{Arkani-Hamed:2023swr}. Let us present an explicit example following \eqref{eq: NLSM example 6pt} at $6$-point:
\begin{equation} 
\eqref{eq: NLSM example 6pt}
\xrightarrow[]{s_{3,6}=0}s_{1,3} \times ( \frac{1}{s_{1,2,3}} +\frac{1}{s_{2,3,4}}   ) \times s_{1,5}
=   {\cal J}^{\mathrm{NLSM}}(3,4,1;\rho^\phi)\;{\cal J}^{\phi^3}(1,\rho^\prime,4,\kappa) {\cal J}^\mathrm{NLSM}(1,4,5;\kappa^\prime).
\end{equation}
For $n=10$, the examples we have shown in section~\ref{sec: splitting of NLSM} can further split into:
\begin{equation}
\eqref{eq: NLSM example 10pt1}
 \xrightarrow[]{m=4} {\cal J}^{\mathrm{NLSM}}(5,2,3; \rho) {\cal J}^\mathrm{\phi^3}(2,\rho^\prime,6,\kappa)\; {\cal J}^{\mathrm{NLSM}}(1,2,5,6,7,8,9; \kappa^{\prime}).
\end{equation}
\begin{equation} 
\eqref{eq: NLSM example 10pt2}
\xrightarrow[]{m=4} {\cal J}^{\mathrm{NLSM}+\phi^3}(5,6^\phi,2^\phi,3; \rho^\phi) {\cal J}^\mathrm{NLSM}(2,\rho^\prime,6,\kappa)\; {\cal J}^{\mathrm{NLSM}+\phi^3}(1,2^\phi,6^\phi,7,8,9; \kappa^{\prime\, \phi}).
\end{equation}

\paragraph{sGal} The factorization near zeros of special Galileon is essentially the same as what we have shown for the NLSM, except that there is no specific ordering; any permutation of such special kinematic would lead to corresponding factorization. Let us present an explicit example following \eqref{eq: sGal example 6pt}:
\begin{equation} 
\begin{aligned}
\eqref{eq: sGal example 6pt}
\xrightarrow[]{s_{3,6}=0}&s_{1,3} s_{3,4} (s_{1,3}+s_{3,4}) \times ( \frac{1}{s_{1,2,3}} +\frac{1}{s_{2,3,4}}   ) \times -s_{1,5} s_{4,5} (s_{1,5}+s_{4,5})\\
=&   {\cal J}^{\mathrm{sGal}}(3,4,1;\rho)\;{\cal J}^{\phi^3}(1,\rho^\prime,4,\kappa) {\cal J}^\mathrm{sGal}(1,4,5;\kappa^\prime).
\end{aligned}
\end{equation}
The 10-point examples at the end of subsubsection~\ref{subsubsec:sGal} also factorizes as:
\begin{equation} 
\eqref{eq: sGal example 10pt1}
     \xrightarrow[]{m=4} {\cal J}^{\mathrm{sGal}}(5,6,3; \rho) {\cal J}^\mathrm{\phi^3}(3,\rho^\prime,6,\kappa)\; {\cal J}^{\mathrm{sGal}}(1,2,5,6,7,8,9; \kappa^{\prime}).
\end{equation}
\begin{equation} 
\eqref{eq: sGal example 10pt2}
     \xrightarrow[]{m=5} {\cal J}^{\mathrm{sGal}+\phi^3}(6,7^\phi,3^\phi,4; \rho^\phi) {\cal J}^\mathrm{sGal}(3,\rho^\prime,7,\kappa)\; {\cal J}^{\mathrm{sGal}+\phi^3}(1,2,3^\phi,7^\phi,8,9; \kappa^{\prime\, \phi}).
\end{equation}

\paragraph{YMS, EMS and DBI} We will demonstrate the details of factorization near zeros for YMS, but a similar procedure holds for EMS and DBI. As mentioned before, it is important to note that the kinematic condition $s_{a,n}=s_{a,\kappa}=0$ with $a\in A\backslash\{m\}$ for the set $\{i,A,j,\kappa\}$ is exactly the condition~\eqref{eq_splitKin} with special legs $i'=j,j'=i,k'=m$. Therefore for odd $|A|$, the additional splitting is applied to the current of pure YMS, and one can easily write the result according to \eqref{eq:YMSAmpcase1}$\sim$\eqref{eq:YMSAmpcase3}. For even $|A|$, one needs to consider the splitting of the mixed current or the object defined via its CHY formula. Quite nicely, the result is universally given by the product of a current of YMS and a 4-point $\phi^3$ one even for the exceptional case~\eqref{eq:YMSAmpcase3}. Concretely, we have:
\begin{equation} 
\begin{aligned}
    &{\cal J}^{\mathrm{YMS}+\phi^3}(\{a_1,a_2\},\ldots,\{i,j,\kappa \})\xrightarrow[]{s_{a,n}=0,a\in A\backslash\{m\}}\\
    & {\cal J}^{\mathrm{YMS}}(\{a_1,a_2\},\ldots,\{\rho,a_r\},\ldots,\{i,j\}) {\cal J}^{\phi^3}(i,\rho',j,\kappa).
\end{aligned}
\end{equation}
for cases correspond to~\eqref{eq:YMSAmpcase1} and~\eqref{eq:YMSAmpcase2} (but note here we assume $|A|$ is even), where $a_r$ refers the particle that is paired with $m$ before further splitting. And similarly, the case corresponds to~\eqref{eq:YMSAmpcase3} is:
\begin{equation}
\begin{aligned}
    &\int d \mu^\mathrm{CHY}_{|A|+3} \Delta_{i,j,\kappa} \frac{\mathrm{Pf} \mathbf{A}_{(i,j)}\mathrm{PT}(i,(i,j),j,\kappa)}{z_{i,a_1}z_{j,a_2}z_{i,\kappa}z_{j,\kappa}\prod_{p}z_{a_p,a_{p+1}}} \xrightarrow[]{s_{a,n}=0,a\in A\backslash\{m\}} \\
    & {\cal J}^{\mathrm{YMS}}(\{i,a_1\},\{j,a_2\},\ldots,\{\rho,a_r\},\ldots,\{a_{|A|-1},a_{|A|}\}) {\cal J}^{\phi^3}(i,\rho',j,\kappa).
\end{aligned}
\end{equation}
Let us present the factorization near zeros of the $10$-point examples considered in ~\ref{sec:YMSfac}:
\begin{equation}
    \eqref{eq: YMS example 10pt1} \xrightarrow[]{m=4} 
     \mathcal{J}^{ \mathrm{YMS}+\phi^3}(\{5,6\},\{4,7,\rho\}) \mathcal{J}^{\rm YMS}(\{3,\rho'\},\{7,\kappa\}) \mathcal{J}^{ \mathrm{YMS}+\phi^3}(\{7,8\},\{1,2\},\{3,9,\kappa'\}).
\end{equation}
\begin{equation}
     \eqref{eq: YMS example 10pt2}  \xrightarrow[]{m=5} 
      \mathcal{J}^{ \mathrm{YMS}}(\{\rho,6\},\{4,7\})  \mathcal{J}^{\phi^3}(4,\rho',j,\kappa) \mathcal{J}^{\rm YMS}(\{1,2\},\{3,4\},\{7,8\},\{9,\kappa'\}). 
\end{equation}

\subsubsection{Factorization near zeros of YM/GR amplitudes}
In exactly the same way, factorizations near zeros for gluons/gravitons follow from their splittings, which we summarize as follows. 
\paragraph{YM} The factorization near zeros of YM amplitudes have two types: one is to further split the current ${\cal J}^{\mathrm{YM}+\phi^3} $; the other is to split the current ${\cal J}^{\mathrm{YM}} $. Both can be achieved by imposing further conditions after the 2-split of YM amplitudes~\eqref{eq:splitting of YM}. Given~\eqref{eq: splitting of PT} and~\eqref{eq: splitting of PfPsi}, it is easy to show the further splitting of the current:
\begin{equation}
   {\cal J}^{\mathrm{YM}+\phi^3} (i^\phi, A, j^\phi; \kappa^\phi)\xrightarrow[a\in A\backslash\{m\}]{p_{a}\cdot p_{n}=\epsilon_{a}\cdot p_{n}=p_{a}\cdot \epsilon_{m}=\epsilon_{a}\cdot \epsilon_{m}=0} {\cal J}^{\mathrm{YM}+\phi^3}(i^\phi, A\backslash\{m\}, j^\phi; \rho^\phi ) {\cal J}_{\mu}^{\mathrm{YM}+\phi^3}(i^\phi,\rho^\prime,j^\phi,\kappa^\phi)\epsilon_{m}^{\mu}\,.
\end{equation}
If we change the 2-split conditions for the YM amplitudes to retain the current $ {\cal J}^{\mathrm{YM}}_{\mu} (i, A, j; \kappa)\epsilon_{n}^{\mu}$, the further conditions are similar to the 2-split conditions\eqref{eq:splitting of YM}, and the current splits as
\begin{equation}
    {\cal J}^{\mathrm{YM}}_{\mu} (i, A, j; \kappa)\epsilon_{n}^{\mu} \xrightarrow[a\in A\backslash\{m\},b\in\{n\},b'\in\{n,i,j,m\} ]{p_a\cdot p_b,\epsilon_a \cdot \epsilon_{b'},p_a \cdot \epsilon_{b'},\epsilon_a \cdot p_b=0} {\cal J}^{\mathrm{YM}+\phi^3}(i^\phi, A\backslash\{m\}, j^\phi; \rho^\phi ) {\cal J}_{\mu\nu}^{\mathrm{YM}}(i,\rho^\prime,j,\kappa)\epsilon_{n}^{\mu} \epsilon_{m}^{\nu}\,.
\end{equation}
where the pure gluon current with two massive legs $\rho^\prime,\kappa$ contract with two polarization vectors $\epsilon_{n}^{\mu},\epsilon_{m}^{\nu}$.

Now, we continue the 7-point example~\eqref{eq_7ptYM2split} to the factorization near zeros. By further splitting either the mixed or the pure current, we can arrive at different splittings.
For the former case, we choose $m=2$ and impose $\epsilon_{3}\cdot \epsilon_{2} = p_{3} \cdot \epsilon_{2} = \epsilon_{3}\cdot p_{7} = p_{3}\cdot p_{7}=0$, such that
\begin{equation}
    \eqref{eq_7ptYM2split} \xrightarrow[]{m=2}   
    {\cal J}^{\mathrm{YM}+\phi^3} (1^\phi,3,4^\phi; \rho^\phi)
    \times {\cal J}^{\mathrm{YM}+\phi^3}_{\nu} (1^\phi,\rho^{\prime},4^\phi, \kappa^\phi) \epsilon^{\nu}_{2}\,
    \times {\cal J}^{\mathrm{YM}}_{\mu} (1,4,5,6; \kappa^{\prime}) \epsilon_{7}^{\mu} .
\end{equation}
For the latter, with $m=5$ and setting $p_6 \cdot p_{7} = \epsilon_{6} \cdot \epsilon_{b^\prime}= p_6 \cdot \epsilon_{b^\prime} = \epsilon_6\cdot p_{7} =0$ for $b^\prime \in \{1,4,5,7\}$, we have
\begin{equation}
    \eqref{eq_7ptYM2split} \xrightarrow[]{m=5}   
    {\cal J}^{\mathrm{YM}+\phi^3} (1^\phi,2,3,4^\phi; \kappa^\phi)\; 
    \times {\cal J}^{\mathrm{YM}}_{\nu\mu} (1,4, \rho^{\prime},\kappa) \epsilon^{\nu}_{5} \epsilon_{7}^{\mu} \,
    \times {\cal J}^{\mathrm{YM}} (6, 1^\phi,4^\phi; \rho^{\phi}).
\end{equation}

\paragraph{GR} For GR amplitudes, the factorization near zeros have more types on the splitting of  ${\cal J}^{\mathrm{GR}}$,  ${\cal J}^{\mathrm{GR}+\phi^3} $, and ${\cal J}^{\mathrm{EYM}} $. But they are very similar to the YM case, so we do not show the details, but just list the conclusions,
\begin{equation}
   {\cal J}^{\mathrm{GR}+\phi^3} (i^\phi, A, j^\phi; \kappa^\phi)\xrightarrow[ \tilde{\epsilon}_{a}\cdot p_{n}=p_{a}\cdot \tilde{\epsilon}_{m}=\tilde{\epsilon}_{a}\cdot\tilde{\epsilon}_{m}=0]{p_{a}\cdot p_{n}=\epsilon_{a}\cdot p_{n}=p_{a}\cdot \epsilon_{m}=\epsilon_{a}\cdot \epsilon_{m}=0}  {\cal J}^{\mathrm{GR}+\phi^3}(i^\phi, A\backslash\{m\}, j^\phi; \rho^\phi ) {\cal J}_{\mu\nu}^{\mathrm{GR}+\phi^3}(i^\phi,\rho^\prime,j^\phi,\kappa^\phi)\epsilon_{m}^{\mu}\tilde{\epsilon}_{m}^{\nu}\,,
\end{equation}
where $a\in A\backslash\{m\}$, and 
\begin{equation}
 {\cal J}^{\rm EYM}_{\nu} (i^g, A, j^g; \kappa^g) \tilde{\epsilon}_n^\nu  \xrightarrow[\tilde{\epsilon}_a \cdot \tilde{\epsilon}_{b'},\tilde{\epsilon}_a \cdot p_b=0]{p_a\cdot p_b,\epsilon_a \cdot \tilde{\epsilon}_{b'},p_a \cdot \tilde{\epsilon}_{b'},\epsilon_a \cdot p_b=0} {\cal J}^{\mathrm{GR}+\phi^3}(i^\phi, A\backslash\{m\}, j^\phi; \rho^\phi ) {\cal J}^{\mathrm{EYM}}_{\nu}(i^g,\rho^\prime,j^g,\kappa^g)\tilde{\epsilon}_n^\nu\,,
\end{equation}
where $a\in A\backslash\{m\},b\in\{n\},b'\in\{n,i,j,m\}$.

For further splitting of ${\cal J}^{\mathrm{GR}}_{\mu\nu} (i, A, j; \kappa)\epsilon_{n}^{\mu}\tilde{\epsilon}_{n}^{\nu}$ the conditions are the same as~\eqref{eq:splitting of GR 1} and \eqref{eq:splitting of GR 2} for $a\in  A\backslash\{m\}$, $b\in \{n\}$, and  $b'\in\{n,i,j,m\}$.

\subsection{Soft limits}

Another implication of even the simplest splitting, which is the special ``skinny" case with $|A|=1$, is the universal behavior when the momentum of the particle in $A$ becomes soft. As we have outlined in~\cite{Cao:2024gln}, such special splitting immediately implies Weinberg's soft theorems for the case of gluon and graviton amplitudes~\cite{Weinberg:1965nx}. Moreover, for Goldstone particles, it implies the (enhanced) Adler zeros for NLSM, DBI and sGal amplitudes~\cite{Adler:1964um,Cheung:2014dqa}. 

Let us begin with ``skinny" splitting of YM amplitude: recall that for the special case $|A|=1$, $n$-gluon amplitude splits into $(n{-}1)$-gluon current and a four-point mixed one; very nicely, this four-point mixed current involving gluon $a$, scalars  $i,j$ as well as an off-shell scalar leg $\kappa$, can be computed exactly even at finite $\alpha'$, in terms of Beta functions:
\begin{equation}\label{eq:mixed 4ptYM}
{\cal J}^{\rm mixed} (i^\phi, a, j^\phi; \kappa^\phi)=\epsilon_a \cdot p_i \, B(\alpha' s_{i,a}, \alpha' s_{j,a}+1) -\epsilon_a \cdot p_j \, B(\alpha' s_{i,a}+1, \alpha' s_{j,a})\,, 
\end{equation}
which in the field-theory limit $\alpha'\to 0$ reduces to an ``off-shell" soft factor $\frac{\epsilon_a \cdot p_i}{s_{i,a}} + (i\to j)$. 

To go from our ``skinny" splitting kinematics to the actual soft limit, we have to be a bit more careful. At this stage we have only imposed $s_{a, b\in B}=0$ where $B=\{1,2,\ldots,n\}\backslash\{a, i, j\}$, which does not imply the softness of $p_a$. The soft limit can be reached by further imposing $s_{a,i}, s_{a,j}\to 0$ (thus $s_{a,n}=0$), in which case the $(n{-}1)$-gluon current becomes an amplitude ($\kappa'$ becomes on-shell). In other words, instead of sending all $s_{a, b'}$ for $b' \in \{1,2,\ldots,n\}\backslash\{a\}$ to zero simultaneously, we are taking a two-step procedure, and we need to ``average" over all possible assignments of $i,j,k \neq a$. For gluon amplitude, we know that $i,j$ must be adjacent to $a$ in the color ordering to contribute at leading order, thus we only need to sum over $k$ where each term gives identical result, thus up to a overall constant we obtain
\begin{equation}
 \sum_{k\neq i,j, a} {\cal J}^{\rm mixed}  \times {\cal J}_{n{-}1}
 \to  \left(\frac{\epsilon_a \cdot p_i}{p_a\cdot p_i}{-} \frac{\epsilon_a \cdot p_j}{p_a\cdot p_j}\right) \times {\cal M}_{n{-}1}^{\rm YM},
\end{equation}
where the mixed current simplifies to nothing but the soft gluon factor, even at finite $\alpha'$! Although we have imposed restrictions on the polarizations~\eqref{eq_pol} not needed for soft limit, they do not appear at leading order, thus we have derived the soft gluon theorem for YM (and bosonic and superstring extensions) and find an interpretation of the soft factor in terms of the four-point mixed current. 

A similar argument applies to the soft graviton, where again we have a four-point mixed current ${\cal J}^{\rm mixed}$ with graviton $a$, scalars $i,j$ and off-shell leg $\kappa$, times the $(n{-}1)$-graviton current. When going to the soft limit with $s_{a, i}, s_{a,j} \to 0$, we need to sum over triplets $i,j,k \neq a$ since mixed current with any choice of $i,j,k$ contributes to the leading soft factor. Up to a overall constant we have
\begin{equation}
\sum_{k,i,j\neq a}{\cal J}^{\rm mixed} \times {\cal J}_{n{-}1}
\to \left(\sum_{b\neq a} \frac{\epsilon_a \cdot p_b \tilde\epsilon_a \cdot p_b}{p_a \cdot p_b}\right)\times {\cal M}_{n{-}1}^{\rm GR} 
\end{equation}
which is the soft graviton theorem (in bosonic and superstring theories) with the soft factor again interpreted from the four-point mixed current. 

Next we move to soft limit of Goldstone scalars in NLSM, DBI and sGal in the field theory limit, where the (enhanced) Alder zeros again immediately follow from corresponding four-point currents. Note that with ``skinny" splitting $|A|=1$, we have (sGal can be replaced by NLSM and DBI):
\begin{equation}
\label{eq_sGal2split}
{\cal A}^{\rm sGal}_n \to {\cal J}^{\rm sGal} (i, a, j; \kappa) \times {\cal J}^{\rm mixed} (i^\phi, \ldots , j^\phi; \kappa'^\phi)\,.  
\end{equation}
Now in the soft limit, we have $p_a=\tau \hat{p}_a$ with $\tau \to 0$, for NLSM, DBI and sGal, the four-point function behaves like ${\cal A}_4 \sim \tau^s$ for $s=1,2,3$, respectively, which immediately leads to the enhanced Adler's zero. What multiplies ${\cal A}_4$ is a $n{-}1$-point mixed current with $3$ $\phi$'s, and at least for NLSM we can start from here and derive the coefficient of Adler zero as summing over such mixed amplitudes with $3$ $\phi$'s, which agrees with the result of~\cite{Cachazo:2016njl}. 

In~\cite{Arkani-Hamed:2024fyd}, multiple soft limits for {\it e.g.} NLSM amplitudes have been studied by considering more general splittings formulated on the surface, and here we also briefly comment on such limits for Goldstone scalars. There are precisely two cases: for $|A|$ odd, the splitting gives a pure current with $|A|+3$ legs and a mixed one, while for $|A|$ even, it gives a mixed current with $|A|+3$ legs (including $3$ $\phi$'s) and a pure one. In the first case, by taking all the momenta in $A$ to be soft, the pure current vanishes just as in the ``skinny" case (again it vanishes at ${\cal O}(\tau)$, ${\cal O}(\tau^2)$ and  ${\cal O}(\tau^3)$ for NLSM, DBI and sGal, respectively), which generalizes the (enhanced) Adler's zeros. In the second case, the simultaneous multi-soft limit then takes the form
\begin{equation}
{\cal A}_n^{\rm sGal} \to {\cal S}^{\rm sGal}_A \times {\cal A}^{\rm sGal}_{n-|A|}(i, B, j, k)\,,\qquad {\cal S}_A=\lim_{p_{a \in A} \to 0} {\cal J}^{\rm mixed}(i^\phi, A, j^\phi; \kappa^\phi)\,,
\end{equation}
where the multi-soft factor ${\cal S}_A$ is given by the mixed currents in the limit where all scalars in $A$ becomes soft (which only depends on the soft momenta $p_{a \in A}$), and very similar results hold for such limits of NLSM, DBI (YMS/EMS) amplitudes as well. For the special case of $|A|=2$, this reduces to the (leading-order) double-soft limits studied for these amplitudes in~\cite{Cachazo:2015ksa}. Of course such results can also be derived from Feynman diagrams, but it is nice to find a direct interpretation of the multi-soft factor in terms of these mixed currents. 

\subsection{Comments on multi-splittings}

Finally, we comment on multi-splittings, where the simplest case is to go from our $2$-split to the so-called ``smooth splitting" or $3$-split of~\cite{Cachazo:2021wsz}. The latter simply follows from further splitting {\it e.g.} $B$ (assume $|B|>1$) as  $B=B'\cup C$ and demand $s_{b,c}=0$ for $b\in B', c\in C$, then the scattering potential splits into three:
\begin{equation}
{\cal S}_n \to {\cal S} (i, A, j; \kappa_A) + {\cal S}(j, B', k; \kappa_B)+{\cal S}(k, C, i; \kappa_C)    
\end{equation}
with off-shell momenta of $\kappa_{A}, \kappa_B, \kappa_C$ given by momentum conservation. As pointed out in~\cite{Cachazo:2021wsz}, such $3$-split really deserves the name ``smooth splitting" since all $n$ on-shell legs appear in the currents (with $i,j,k$ each shared by two of them and the symmetry between $i,j,k$ restored). Our special ``skinny'' case where we have {\it e.g.} $|A|=1$ corresponds to the special case of~\cite{Cachazo:2021wsz} where one of the three currents becomes the trivial $3$-point one. Exactly the same argument applies to the scattering equations/CHY measure, thus we can generalize the smooth splitting for scalar amplitudes~\cite{Cachazo:2021wsz} to gluon/graviton amplitudes (and their string extensions). 

Note that it is not obvious that we could go further by iterating the procedure, since that would require a good understanding of splitting off-shell currents expressed using string/CHY integrals. However, without asking about physical interpretations, it is clear that string/CHY integrals do factorize into lower-point integrals when we go to more and more special kinematics (see~\cite{Arkani-Hamed:2024yvu} for related discussions). To illustrate this point, let us focus on the other extreme when the amplitude ``maximally splits" into the product of $n{-}3$ four-point currents (one-fold string/CHY integrals), which was known as ``minimal kinematics"~\cite{Cachazo:2020uup, Early:2024nvf, Arkani-Hamed:2024yvu}. For (stringy) $\phi^3$ case, it is well known that we need the special kinematics of the form:
\begin{equation}\label{eq:multi-split}
    s_{1,i_{1}} = s_{2,i_{2}} = \ldots = s_{n-4,i_{n-4}} = 0\,,
\end{equation}
where $i_{c} \in \{j+2, \ldots, n-1\}$. Such an maximal split can be regarded as $(n-4)$ iterations of the $2$-split, where \eqref{eq:multi-split} is equivalent to $(n-4)$ $2$-split kinematics. For the first one, we choose $i=2, j=n, k=3$, leading to the split kinematics $s_{1,i_{1}}=0$. Then the amplitude splits into a four-point current and an $(n-1)$-point current. For further 2-split of the $(n-1)$-point current, we can choose $i=3, j=n, k=4$, with the condition $s_{2,i_{2}}=0$. By recursively applying further 2-split, we eventually achieve the maximal-split.
\begin{equation}
    \mathcal{M}_{n}^{\phi^3} \xrightarrow[]{\eqref{eq:multi-split}} \prod_{i=3}^{n{-}2}B(\alpha' X_{1,i}, \alpha' X_{i,n})\,.
\end{equation}
where  $X_{i,j} = (p_i + \dots +p_{j{-}1})^2$. In the field-theory limit $\alpha'\to 0$, the Beta function $B(\alpha' X_{1,i}, \alpha' X_{i,n})$ reduces to $\left(\frac{1}{X_{1,i}} + \frac{1}{X_{i,n}}\right)$.

As an example, for the $7$-point  $\phi^3$ amplitude, the minimal kinematics are $s_{1,4}=s_{1,5}=s_{1,6}=s_{2,5}=s_{2,6}=s_{3,6}=0$, the amplitude splits as,
\begin{equation}
    \mathcal{A}_{7}^{\phi^3} \xrightarrow[]{\eqref{eq:multi-split}} \left(\frac{1}{s_{1,2}} + \frac{1}{s_{1,7}}\right) \left(\frac{1}{s_{1,2,3}} + \frac{1}{s_{1,2,7}}\right) \left(\frac{1}{s_{5,6,7}} + \frac{1}{s_{4,5,6}}\right) \left(\frac{1}{s_{6,7}} + \frac{1}{s_{5,6}}\right)\,.
\end{equation}

For the maximal-splittings of gluon amplitudes, the maximal-split conditions should include the constraints like \eqref{eq_pol}. Specifically, these conditions are 
\begin{equation}\label{eq:multi-split-pol}
    \epsilon_{1} \cdot \epsilon_{i_{1}^{\prime}} = p_{1} \cdot \epsilon_{i_{1}^{\prime}} = \epsilon_{1} \cdot p_{i_{1}} = \ldots = \epsilon_{n-3} \cdot \epsilon_{i_{n-3}^{\prime}} = p_{n-3} \cdot \epsilon_{i_{n-3}^{\prime}} = \epsilon_{n-3} \cdot p_{i_{n-3}} = 0\,,
\end{equation}
where $i_{c} \in \{j+2, \ldots, n-1\}$ and $i_{c}^{\prime} \in \{j, \ldots, n\}$. The important aspect of maximal-splittings of gluon amplitudes is that the $n$-point amplitude splits into $(n-3)$ four-point mixed currents (1 gluon and 3 $\phi$'s) and a three-point pure gluon current. The reason is that the four-point pure gluon amplitude can split into a four-point mixed current and a three-point pure gluon current~\eqref{eq:ym4pt2split}. The gluon string amplitude splits as,
\begin{equation}
    \mathcal{M}_{n}^{\rm YM} \xrightarrow[]{\eqref{eq:multi-split-pol}} {\cal J}(n{-}2,n{-}1,n)\prod_{i=1}^{n{-}3}{\cal J}^{\rm mixed}(i,(i{+}1)^{\phi},n^{\phi},\kappa_i^{\phi})\,.
\end{equation}
where the pure gluon three-point current is the same as the three-point amplitude, and the mixed current is the same as~\eqref{eq:mixed 4ptYM}, $p_{\kappa_{i}}=-p_{i}-p_{i{+}1}-p_{n}$. In the field-theory limit $\alpha'\to 0$, the mixed current ${\cal J}^{\rm mixed}(i,(i{+}1)^{\phi},n^{\phi},\kappa_i^{\phi})$ reduces to the four-point current ${\cal J}^{{\rm YM}+\phi^3} (i,(i{+}1)^{\phi},n^{\phi},\kappa_i^{\phi})$.

\section{Conclusions and Outlook}
In this paper, we have presented a proof of the newly discovered splitting behavior for a large class of tree-level amplitudes expressed either as string integrals or CHY formulas, which is based on a detailed study of such behavior for various string correlators and CHY integrands. This has provided a worldsheet origin for the splitting behavior universally present in all these string and particle amplitudes, which in turn explains other types of new behavior of tree amplitudes such as smooth splitting (and multi-splittings) of~\cite{Cachazo:2021wsz}, as well as factorizations near zeros~\cite{Arkani-Hamed:2023swr}. Among other things, we have generalized the former to string amplitudes, the latter to amplitudes without color, and both of them to amplitudes of gluons and gravitons in YM/GR as well as bosonic/superstring theory, thus putting all these new ``factorizations" on a firm ground for tree amplitudes in a web of theories. We emphasize that all these ``factorization" behavior (and consequently zeros of amplitudes) are otherwise difficult to see in the conventional formulation of QFT, and they have numerous interesting physics implications, {\it e.g.} Weinberg's soft theorems for gluons/gravitons and (enhanced) Adler's zeros for Goldstone particles already follow from the special ``skinny" splitting case. 

Our results have opened up several directions for future investigations. First of all, an interesting question one can ask is if the splitting can be explained purely in field-theory context without referring to the worldsheet. In particular, we have seen that the splitting produces mixed currents and in certain cases (such as mixed currents of Yang-Mills-scalar theory) we have defined the mixed currents purely in terms of string/CHY formulas where the precise Lagrangian or a clear physics picture is still lacking; it would be highly desirable to understand precisely what are these mixed currents/amplitudes without referring to the worldsheet. A related question is if tree amplitudes in more general theories (such as those with higher-dimensional operators) also have such splitting behavior, and in this way if one could determine the ``landscape" or web of theories that all split nicely. One could imagine that maybe such splitting behavior is actually related to some hidden symmetry or universal properties of these amplitudes (like color-kinematics duality and double copy~\cite{Bern:2008qj, Bern:2019prr}). For example, could we derive all the splitting behavior directly from the web of relations among all these amplitudes~\cite{Cachazo:2014xea, Cheung:2017ems, Dong:2021qai} and/or their BCJ numerators~\cite{Du:2017kpo, Edison:2020ehu, He:2021lro}?

Relatedly, we have seen that already the special splitting is closely related to soft limit: for gluon and graviton it implies Weinberg's soft theorem and for Goldstone particles it clearly leads to (enhanced) Adler's zero for NLSM, DBI and sGal. It would be interesting to see if the splitting could account for subleading corrections to the soft theorems~\cite{Low:1958sn,Cachazo:2014fwa}. For Adler's zero, our results indicate that the coefficient of such zeros must be related to the mixed amplitude with $3$ $\phi$'s as studied in~\cite{Cachazo:2016njl}. It would be highly desirable to derive the result of~\cite{Cachazo:2016njl} systematically from splitting and ask if one could generalize the argument to multi-particle soft limits (for double-soft limits, subleading corrections are studied in~\cite{Cachazo:2015ksa}). Another important direction would be to utilize such splitting behavior (or even just the zeros) to learn more about all these string and particle amplitudes, {\it e.g.} could we fully determine these tree amplitudes from their splitting behaviors?

Moreover, we have some remarks regarding zeros and factorizations near zeros for YM and gravity amplitudes. Unlike scalar cases, here we need additional conditions on polarizations already for the splitting, and even more so for (factorizations near) zeros. We certainly do not claim to exhaust all possible zeros and factorizations for these amplitudes. For example, even without touching any Mandelstam variables, YM and gravity amplitudes trivially vanish when we set $\epsilon_a \cdot k_b=\epsilon_a \cdot \epsilon_b$ for a given $a$ and all $b\neq a$, and similarly gauge invariance can also be viewed as certain kind of zero conditions (see~\cite{Arkani-Hamed:2024fyd} for a unified form for both properties). It would be nice to carve out the space of most general zeros of YM/gravity amplitudes (and possible factorizations near them). On the other hand, one can also ask about zeros of helicity amplitudes in four dimensions, which can involve very different types of conditions than their $D$-dimensional counterpart. An obvious class of four-dimensional zeros seem to be setting $\langle i,j\rangle=0$ for all $i,j$ being negative-helicity gluon/graviton or the parity conjugate ($[i,j]=0$ for positive-helicity ones), which makes every BCFW term vanish identically. It would be very interesting to explore this direction further. 

Last but not least, all we have considered so far are for tree-level amplitudes, but very recently in~\cite{Arkani-Hamed:2024fyd} the splitting for (stringy) Tr $\phi^3$ loop integrands has been understood as coming from gluing smaller surfaces; by deforming the (stringy) Tr $\phi^3$ to NLSM or $2n$-scalar YMS~\cite{Arkani-Hamed:2023swr}, it is expected that such splitting also applies to the loop integrands for those cases. It would be extremely interesting if we could understand such splittings of loop integrands for more general amplitudes involving Goldstone particles, gluons and gravitons. At least for one-loop case, we can also hope to derive such splitting behavior for GR and DBI/sGal {\it etc.} from either one-loop CHY formulas~\cite{Geyer:2015bja, Cachazo:2015aol, He:2015yua} or combining one-loop double copy(c.f.~\cite{He:2017spx, Edison:2020uzf, Edison:2022jln}) with splitting of NLSM and YMS amplitudes~\cite{Arkani-Hamed:2024fyd}. We would like to further study such splitting behaviors and (factorizations near) zeros for loop integrands and implications for loop amplitudes from all these different perspectives. Given that Adler's zero can be derived from the ``skinny" splitting case, it would be interesting to see how Adler's zero of loop integrands in NLSM~\cite{Arkani-Hamed:2024nhp, Bartsch:2024ofb} (and the ``surface zero" of~\cite{Arkani-Hamed:2024fyd}) as well as DBI/sGal may follow from ``skinny" splitting at loop level, and how to understand soft theorems of gluons/gravitons at loop level (c.f.~\cite{Bern:2014oka, He:2014bga}) from this point of view. 

\paragraph{Acknowledgments}
It is our pleasure to thank Nima Arkani-Hamed, Carolina Figueiredo for inspiring discussions and collaboration on related projects, and Freddy Cachazo, Jaroslav Trnka, Laurentiu Rodina and Yong Zhang for communications regarding related works. The work of SH is supported by the National Natural Science Foundation of China under Grant No. 12225510, 11935013, 12047503, 12247103, and by the New Cornerstone Science Foundation through the XPLORER PRIZE. The work of CS is supported by China Postdoctoral Science Foundation under Grant No. 2022TQ0346, and the National Natural Science Foundation of China under Grant No. 12347146.

\bibliographystyle{JHEP}
\bibliography{fac}
\end{document}